\documentclass[11pt]{article}

\usepackage{amsmath}
\usepackage{amssymb}
\usepackage{setspace}

\usepackage{graphicx}

\usepackage{cite}

\usepackage{hyperref}

\usepackage{lineno}

\usepackage{microtype}
\usepackage{color}
\usepackage{lineno}
\DisableLigatures[f]{encoding = *, family = * }

\usepackage[labelfont=bf,labelsep=period,justification=raggedright]{caption}

\makeatletter
\renewcommand{\@biblabel}[1]{\quad#1.}
\makeatother

\date{}
\begin{document}

% Title must be 150 characters or less
\begin{flushleft}
{\Large
	\textbf{Training and spontaneous reinforcement of neuronal assemblies by spike timing}
}
% Insert Author names, affiliations and corresponding author email.
\\
Gabriel Koch Ocker$^{1,3,4}$, 
Brent Doiron$^{2,3}$
\\
\bf{1:} Department of Neuroscience, University of Pittsburgh, Pittsburgh, PA, USA
\\
\bf{2:} Department of Mathematics, University of Pittsburgh, Pittsburgh, PA, USA
\\
\bf{3:} Center for the Neural Basis of Cognition, University of Pittsburgh and Carnegie Mellon University, Pittsburgh, PA, USA
\\
\bf{4:} Allen Institute for Brain Science, Seattle, WA, USA

\end{flushleft}

% Please keep the abstract between 250 and 300 words
\section{Abstract}
\paragraph{}
The synaptic connectivity of cortex is plastic, with experience shaping the ongoing interactions between neurons. Theoretical studies of spike timing--dependent plasticity (STDP) have focused on either just pairs of neurons or large-scale simulations where analytic insight is lacking.   A simple account for how fast spike time correlations affect both micro- and macroscopic network structure remains lacking. We develop a low-dimensional mean field theory showing how STDP gives rise to strongly coupled assemblies of neurons with shared stimulus preferences, with the connectivity actively reinforced by spike train correlations during spontaneous dynamics. Furthermore, the stimulus coding by cell assemblies is actively maintained by these internally generated spiking correlations, suggesting a new role for noise correlations in neural coding.  Assembly formation has been often associated with firing rate-based plasticity schemes; our theory provides an alternative and complementary framework, where temporal correlations and STDP form and actively maintain learned structure in cortical networks.

\section{Introduction}
%Structure in cortical networks and relation to stimulus preferences.  spike timing--dependent plasticity and precise spike time correlations.  Recent work: experimental results on rates, modeling work relating spike-based to rate-based rules, assemblies in rate models.  Potential role of spike timing?

\paragraph{}
A cornerstone principle that bridges systems and cellular neuroscience is that the synaptic wiring between neurons is sculpted by experience.  The early origins of this idea are often attributed to Donald Hebb \cite{hebb_organization_1949,markram_history_2011,harris_cortical_2013}, who famously postulated that groups of neurons that are repeatedly coactivated will strengthen the synaptic wiring between one another.  The interconnected group, termed an {\it assembly}, has become an essential building block of many theories of neural computation \cite{buzsaki_neural_2010} and associative memory \cite{neves_synaptic_2008}.  Despite the functional appeal of neuronal assemblies, only recently has physiological evidence of assembly structure been collected.            

\paragraph{}
In mouse primary visual cortex, new advances in recording techniques have shown that pyramidal neurons with similar stimulus preferences connect more frequently, with more synapses and with stronger postsynaptic potentials than neurons with dissimilar stimulus preferences \cite{ko_functional_2011, cossell_functional_2015, lee_anatomy_2016}. %Similar trends hold with respect to reciprocal connectivity between pairs of similarly or dissimilarly tuned neurons \cite{cossell_functional_2015}.  
Synaptically connected neurons tend to receive more common inputs than would be expected by chance, suggesting a clustered architecture \cite{yoshimura_excitatory_2005, perin_synaptic_2011}. While strong recurrent connectivity between similarly tuned neurons is present even at eye opening, it is enhanced during development and especially by visual experience \cite{ko_emergence_2013, ko_emergence_2014}.  This suggests long-term synaptic plasticity as a key mechanism for the assembly organization of cortical circuits.  However, we have only a partial understanding about how the mechanics of synaptic plasticity interacts with recurrent circuits to support the training and maintenance of assembly structure. 

\paragraph{}
Physiological investigation over the past two decades has uncovered spike timing--dependent plasticity (STDP) mechanisms whereby the temporal correlations of pre- and postsynaptic spiking activity drive learning \cite{markram_spike-timing-dependent_2012}.  Hebbian STDP reinforces temporally causal interactions between neurons: the connections from presynaptic neurons that causally contribute to a postsynaptic neuron's firing are strengthened, while the other connections are weakened.  Consequently, many modeling studies show that Hebbian STDP promotes the development of feedforward networks \cite{masuda_formation_2007, takahashi_self-organization_2009, tannenbaum2016shaping} with temporally precise \cite{gerstner_neuronal_1996} and tuned \cite{song_cortical_2001} responses giving rise to sequential activity \cite{gerstner_why_1993, fiete_spike-time-dependent_2010}.   Feedforward structures are quite distinct from the recurrent wiring within neuronal assemblies and thus it is not obvious that STDP will support assembly formation.  Nevertheless, recent theoretical work has shown that networks of recurrently coupled spiking neurons having STDP in excitatory connections effectively learn assembly structure \cite{mongillo_learning_2005,litwin-kumar_formation_2014, zenke_diverse_2015} that is stable in the face of ongoing spontaneous spiking activity post training \cite{litwin-kumar_formation_2014, zenke_diverse_2015}.

\begin{figure}[ht!]
\centering \includegraphics{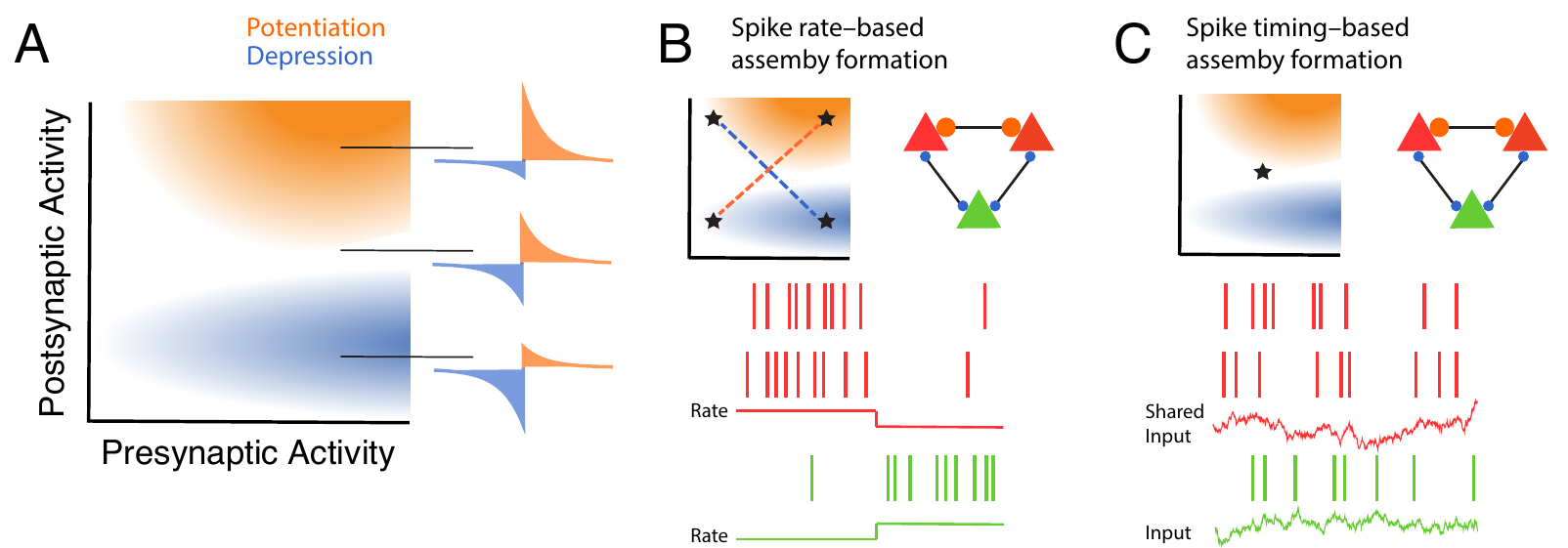}
\caption[{\bf \tiny Balance between potentiation and depression in different plasticity models.}]{
\footnotesize
{\bf Spike rate-- versus spike timing--based neuronal assembly formation.}  (A). Schematic illustrating how combinations of pre- and postsynaptic activity combine to drive synaptic potentiation and depression in models of STDP.  The schematic is adapted from Litwin-Kumar \& Doiron \cite{litwin-kumar_formation_2014} where STDP rules based on third-order spike interactions \cite{pfister_triplets_2006}, or voltage-\cite{clopath_claudia_connectivity_2010} or calcium-based \cite{graupner_calcium-based_2012} learning were studied.  Pre- and postsynaptic correlations are neglected.  The STDP curves on the right indicate the degree of potentiation and depression as pre- and postsynaptic activity ranges.  (B) Example three neuron group with two neurons having co-fluctuating firing rates (bottom, red--red) and the other neuron having anti-correlated firing rate fluctuations (bottom, green--red).  This dynamic potentiates synaptic coupling between correlated neurons while depressing synaptic coupling between anti-correlated neurons (right graph).    (C) Same as B except firing rates are fixed at a value that balances rate-based potentiation and depression.  Shared input correlations to two neurons can potentiate strong recurrent synapses (bottom, red-red) and depress uncorrelated neurons (bottom, green-red). 
}  
\label{Fig_schematic}
\end{figure}

\paragraph{}
The synaptic plasticity models  \cite{pfister_triplets_2006, clopath_voltage_2010, graupner_calcium-based_2012} used in these studies \cite{litwin-kumar_formation_2014, zenke_diverse_2015} capture the known firing rate dependence of the balance between potentiation and depression \cite{sjostrom_rate_2001}.   When spike time correlations are neglected, these models admit reductions of STDP learning to more classic rate-based plasticity schemes \cite{pfister_triplets_2006, gilson_emergence_2009-1, clopath_claudia_connectivity_2010} so that when high (low) postsynaptic activity is paired with high presynaptic activity, synaptic connections are potentiated (depressed) (Figure \ref{Fig_schematic}A).  In these models, the assembly structure in recurrent networks forms via firing rate transitions that toggle between strongly potentiation- and depression-dominated regimes (Figure \ref{Fig_schematic}B).  While these past studies \cite{litwin-kumar_formation_2014, zenke_diverse_2015} show that assembly formation can co-occur with STDP, in these networks any fast spike time correlations between neurons contribute minimally to synaptic learning.

%%%%%%%%%%%%%%%%%%%%%%%%%%%%%%%%%%%%%%%%%%%%%%%%%%%

%%%%%%%%%%%%%%%%%%%%%%%%%%%%%%%%%%%%%%%%%%%%%%%%%

\paragraph{}
Spike trains in diverse cortical areas do exhibit covariable trial-by-trial fluctuations (noise correlations).  These noise correlations covary with neurons' stimulus preferences (signal correlations) \cite{bair_correlated_2001, kohn_stimulus_2005, rothschild_functional_2010} and synaptically connected neurons have higher noise correlations \cite{ko_functional_2011, cossell_functional_2015}, suggesting that assembly structure and noise correlations are related.  Furthermore, excitatory-inhibitory interactions within cortical circuits create nearly synchronous ($\sim 10$ ms) joint temporal structure between spike trains that overlaps with the fine timescale required for STDP learning \cite{borgers_minimal_2012,jia_gamma_2013,salkoff_synaptic_2015}.
In complementary research, several in vivo studies show that the precise timing of pre- and postsynaptic spikes can be a crucial determinant of plasticity \cite{feldman_spike-timing_2012}.
%\cite{pawlak_invivo_2013, meliza_receptive_2006,jacob_invivo_2007,nishimura_spike-timing-dependent_2013,conde_reversed_2013,koch_hebbian_2013}.  
In particular, Kim et al. \cite{kim_emergence_2016} have recently shown that correlations on the order of tens of milliseconds control assembly formation in vivo.  Thus, while previous modeling studies did not require fast spike train correlations to train stable assembly structure \cite{litwin-kumar_formation_2014, zenke_diverse_2015}, there is sufficient experimental evidence to suggest that spike timing may nonetheless play an important role in assembly formation and stability.    

\paragraph{}
Here, we show that spike time correlations can, in the absence of rate-based plasticity mechanisms, form Hebbian assemblies in response to correlated external inputs to groups of neurons (Figure \ref{Fig_schematic}C).  We extend past studies \cite{ocker_self-organization_2015, tannenbaum2016shaping, gilson_emergence_2009-1} and combine linear response theory for spike train covariances in recurrent networks \cite{trousdale_impact_2012} with a slow-fast theory of STDP \cite{kempter_hebbian_1999} to develop low-dimensional theories describing the evolution of the network structure.  Our analyses reveal that training promotes strong connectivity and strong \emph{reciprocal} connectivity within co-stimulated groups.  We further show that after training and in the absence of any external input correlations, internally generated spike time correlations reinforce learned architectures during spontaneous activity.  Finally, this result motivates us to speculate on a new beneficial role of internally generated noise correlations on stimulus coding: to maintain stimulus-specific assembly wiring that supports enhanced response sensitivity.  In total, our theory reveals a potential role for precise spike time correlations in the formation of neuronal assemblies in response to correlated external inputs, as well as their active maintenance during spontaneous activity.  

\section{Results}
\subsection{Plasticity of partially symmetric networks during spontaneous activity}

\paragraph{}
We first present the basic network properties of our network (see Methods \ref{sec:network}). 
 One striking feature of cortical networks is the overrepresentation of reciprocally connected pairs of excitatory neurons, compared to a simple randomly wired (Erd\H{o}s-R\'enyi) network \cite{song_highly_2005, perin_synaptic_2011}. In order to reflect this structure, we took the baseline excitatory-excitatory connectivity of our network, $\mathbf{W}^0_{EE}$, to be composed of two parts: $\mathbf{W}^0_{EE} = \mathbf{W}^0_\text{sym} + \mathbf{W}^0_\text{asym}$, where $\mathbf{W}^0_\text{sym}$ is a symmetric random binary matrix with connection probability $\Omega p_0$ and $\mathbf{W}^0_\text{asym}$ a random binary matrix with connection probability $(1-\Omega)p_0$ (without any symmetry constraint).  Both had Erd\H{o}s-R\'enyi statistics.  The parameter $\Omega$ thus determined the frequency of bidirectionally connected pairs of excitatory neurons in $\mathbf{W}^0_{EE}$.  We modeled networks of 1500 excitatory neurons and 300 inhibitory neurons, both types following exponential integrate-and-fire dynamics \cite{fourcaud-trocme_how_2003}.  The overall connection probability between excitatory neurons was $p_0 = 0.15$, with $\Omega = 0.4$.  Excitatory-inhibitory, inhibitory-excitatory and inhibitory-inhibitory connectivity were asymmetric ($\Omega=0$), with connection probability $0.4$. 

\paragraph{} 

Before we proceed to the formation of assembly structure, we present the underlying synaptic dynamics of both the excitatory-excitatory and inhibitory-excitatory connections in the network in the absence of a training signal. 

\subsubsection{Excitatory plasticity and thresholds for synaptic weight dynamics }
\begin{figure}[ht!]
%\begin{minipage}[c]{0.6\linewidth}
\centering \includegraphics{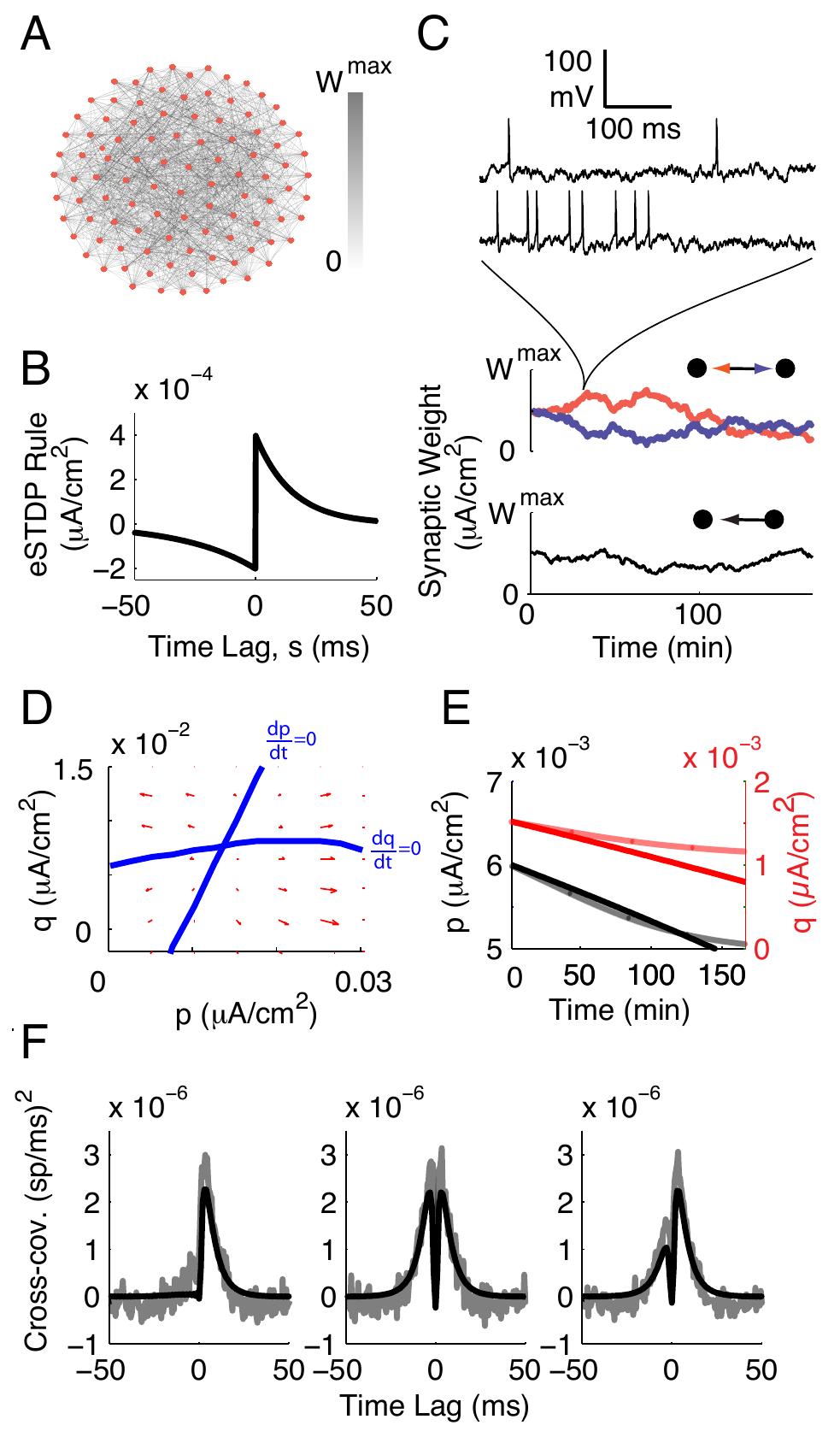}
%\end{minipage}
\hspace{.05\linewidth}
%\begin{minipage}[c]{0.3\linewidth}
\centering \caption[{\bf \tiny Network structure shapes synaptic plasticity.}]{
\footnotesize
{\bf Network structure shapes synaptic plasticity.} (A) Visualization of a random subset of the excitatory-excitatory connectivity.  (B) The eSTDP rule, $L(s)$, is composed of exponential windows for depression (-) and potentiation (+).  Each is defined by its amplitude $f_\pm$ and timescale $\tau_\pm$.  (C) Synaptic weights evolve on a slow timescale.  Individual synaptic weights are governed by the relative timing of spikes in the pre- and postsynaptic neurons' spike trains.  (D) Dynamics of the mean synaptic weight ($p$) and the mean above-chance strength of \emph{reciprocal} synapses, $q$.  There is a threshold for potentiation of each given by its nullcline (blue lines).  (E) Time course of $p$ and $q$ in the case where both are depressing.  Solid lines: theory, Eqs. \eqref{eq:dpEEdt},\eqref{eq:dqEEdt}.  Shaded lines: simulation of the spiking network.  (F) Average spike train covariance between monosynaptically connected pairs (left), reciprocally connected pairs (right) and all pairs (right).  Shaded lines: simulation.  Solid lines: linear response theory (first-order truncation, Eq. \eqref{Ctrunc_inh}).} 
\label{Fig_assembly1}
%\end{minipage}
\end{figure}

\paragraph{}
In this study we consider the evolution of the weighted connectivity $\mathbf{W}_{EE}$ (Figure \ref{Fig_assembly1}A). %; we began by studying plasticity in these networks during spontaneous activity.  
In order to focus on learning due to precise spike time correlations, we used a classical Hebbian spike pair--based plasticity rule for the plasticity between excitatory neurons (eSTDP) \cite{gerstner_neuronal_1996, markram_regulation_1997, bi_synaptic_1998} (Figure \ref{Fig_assembly1}B).  The plasticity rule is phenomenological, and embodies the simple observation that spike pairs induce changes in synaptic weights and the amplitude of these changes depends on the time lag between the two spikes \cite{markram_spike-timing-dependent_2012}.  The coupling strength scaled with system size as $\epsilon = (Np_0)^{-1}$ so that for large $N$ the evolution of $\mathbf{W}_{EE}$ was slow compared to the fast timescales of membrane dynamics and spike discharge (Figure \ref{Fig_assembly1}C).  The separation of timescales between spike time and synaptic weight dynamics permitted an averaging theory for the joint dynamics of $\mathbf{W}_{EE}$ and the spike time covariance  $\mathbf{C}(s)$ (see \cite{ocker_self-organization_2015} for a full description).     

%\paragraph{}
%When individual pairs of pre- and postsynaptic spikes induce small changes in the synaptic weights, those weights evolve on a slow timescale compared to the window of the STDP rule.  This allows their dynamics to be written as a drift-diffusion process \cite{kempter_hebbian_1999}, and the drift of the weights can be written as \ref{sec:plast}:
%\begin{equation}
%\frac{d\mathbf{W}_{ij}}{dt} = \mathbf{W}^0_{ij} \int_{-\infty}^\infty L(s) \left(r_ir_j + \mathbf{C}_{ij}(s) \right) ds
%\end{equation}
%These dynamics embody the simple observation that when spike pairs induce changes in synaptic weights, and the amplitude of these changes depends on the time lag between the two spikes, then the frequency of spike pairs at different time lags interacts with the learning rule to determine the evolution of synaptic weight.  Fortunately, techniques exist to calculate the spike train covariability of pairs of neurons in recurrent networks.  Using a linear response formalism for calculating the spike train covariances $\mathbf{C}(s)$, we approximate them with the contribution of length one paths through the network as in  the previous chapter.  This allows us to develop closed, low-dimensional theories for the plasticity of \emph{statistics} of the network structure.

\paragraph{}
We began with a simple characterization of the network excitatory-excitatory structure in terms of two variables:
\begin{equation} \begin{aligned}
\epsilon p &= \frac{1}{N_E^2} \sum_{i,j \in E} \mathbf{W}_{ij} \\
\epsilon q &= \frac{1}{N_E^2} \sum_{i,j \in E} \mathbf{W}^0_{ij} \mathbf{W}_{ji} -  \epsilon p_0 p
\end{aligned} \end{equation}
These measure the mean weight of excitatory-excitatory synapses ($p$) and the mean weight of \emph{reciprocal} excitatory-excitatory synapses ($q$) above what would be expected in an unstructured network.  (Here, $q$ corresponds to $q_\mathrm{X}^\mathrm{rec}$ in \cite{ocker_self-organization_2015}). Note that with asymmetric connectivity, $\Omega = 0$, $q$ becomes weak $(\mathcal{O}(N^{-3/2}))$ so that the network connectivity can be described (to leading order) only by $p$.  %This corresponds to the unstructured invariant set of \cite{ocker_self-organization_2015}.  
The structure we impose on the network by setting $\Omega \neq 0$ enforces that the variables $p,q$ form, to leading order, an invariant set for the plasticity of synaptic motifs \cite{ocker_self-organization_2015}.

\paragraph{}
We derived dynamics for these variables following the same steps as in \cite{ocker_self-organization_2015} (see Methods, \ref{sec:deriv}).  We first approximated the average spike train covariance from the contributions of length one paths in the network and neglected the bounds on synaptic weights in the eSTDP rule, so that this theory does not account for equilibrium states of the weights.  The network structure $p,q$ then obeys:
\begin{equation}  \label{eq:dpEEdt}
\frac{dp}{dt} = \left(r_E^2S + c_{EE}\sigma^2 S_\eta\right)p_0 + \epsilon \left[S_Fp + S_B\left(q + p_0p\right) + S_C p^2 + S_C^I\gamma (p_{EI}^*)^2\right] \\
\end{equation}
\begin{equation}  \label{eq:dqEEdt}
\frac{dq}{dt} = \left(r_E^2S + c_{EE}\sigma^2 S_\eta\right)q_0 + \epsilon \left[S_Fq + S_B\left(1-p_0\right)\left(q + p_0p\right) + S_C \frac{q_0}{p_0}p^2 + S_C^I \gamma \frac{q_0}{p_0} (p_{EI}^*)^2\right]
\end{equation}
The first terms on the right-hand side of Eq. \eqref{eq:dpEEdt} describe the contributions of chance spike coincidences ($r_E^2 S$), with $r_E$ being the network-averaged firing rate, and correlations induced by external inputs ($c_{EE}\sigma^2S_\eta$).  $S$ is the integral of the eSTDP rule, while $S_\eta$ is the integral of the eSTDP rule against the average susceptibility of two neurons to externally induced correlations (Methods, \ref{sec:deriv}).  The latter terms describe the contribution of correlations induced by coupling within the network, weighted by the eSTDP rule.  The effect of correlations due to direct (forward) connections is measured by $S_F$, and those due to reciprocal (backward) connections is measured by $S_B$.  The final terms arise from correlations due to common inputs from excitatory ($S_C$) or inhibitory ($S_C^I$) neurons.  The parameter $\gamma$ is the ratio of the number of inhibitory neurons to excitatory neurons (here $\gamma=1/3$) and we defer a treatment of the inhibitory to excitatory connection strength $p_{EI}^*$ until the next section.   Finally, $q_0$ is the empirical frequency of reciprocal synapses in the network above chance levels, analogous to $q$ but measured from the adjacency matrix rather than the weight matrix.

\paragraph{}
We took there to be a balance between potentiation and depression, so that $S\sim \mathcal{O}(\epsilon)$ (star in Figure \ref{Fig_schematic}C), with that balance tilted slightly in favor of depression (so that $S < 0$).  This assumption, when combined with an absence of training ($c_{EE}=0$), leads to the synaptic dynamics being governed by different sources of internally generated spiking covariability, each interacting with the eSTDP rule $L(s)$.  Spiking covariations from direct connections mainly contribute at positive time lags, interacting with the potentiation side of the eSTDP rule.  This is reflected in the average spike train covariance between monosynaptically connected neurons (Figure \ref{Fig_assembly1}F, left).  Reciprocal connections, in contrast, contribute spiking covariations at negative time lags, interacting with the depression side of the eSTDP rule.  This is reflected in the average spike train covariance between reciprocally connected pairs, which includes the contributions from both direct and reciprocal connections (Figure \ref{Fig_assembly1}F, middle).  Finally, the contributions from common inputs are temporally symmetric around zero time lag, interacting with both the potentiation and depression windows. The average spike train covariance between all neurons was asymmetric because of the higher frequency of monosynaptically connected over reciprocally connected neurons (Figure \ref{Fig_assembly1}F, right).

%%%%%%%%%%%%%%%%%%%%%%%%%%%%%%%%%%%%%%%%%%%%%%%%%

%%%%%%%%%%%%%%%%%%%%%%%%%%%%%%%%%%%%%%%%%%%%%%%%%

\paragraph{}
The competition between these sources of spiking covariability imposed thresholds for potentiation and depression of the mean field variables $p$ and $q$.  This is best understood by examining the $p$ and $q$ nullclines (Figure \ref{Fig_assembly1}D, blue lines).  Briefly, a nullcline is the collection of $(p,q)$ values where (for example) $dp/dt=0$; on either side of the nullcline the sign of $dp/dt$ dictates the evolution for $p$.  The nullclines of Eqs. \eqref{eq:dpEEdt} and \eqref{eq:dqEEdt} intersect at a single point in $(p,q)$ space, and for the Hebbian plasticity rule used (Figure \ref{Fig_assembly1}B) that point was an unstable repeller, with dynamics flowing away from the point (Figure \ref{Fig_assembly1}D, red arrows).  In this case the nullclines then acted as thresholds so that if either $p$ or $q$ were initially stronger than its threshold it would potentiate, and otherwise it would depress.  It has been long known that additive Hebbian eSTDP produces unstable synaptic dynamics for pairs of coupled neurons through a competition between potentiation and depression \cite{song_competitive_2000,vanRossum_stable_2000}. Our theory has extended this idea to large populations of neurons through mean field dynamics of $p$ and $q$.  

\paragraph{}

Our theory not only gives a qualitative understanding of synaptic dynamics, it also provides a good quantitative prediction of the plasticity within our large-scale integrate-and-fire network (Figure \ref{Fig_assembly1}E, compare the solid theory curves to the shaded curves estimated from numerical simulations).  The threshold dynamics for $p$ and $q$, and their dependence upon various aspects of spike time correlations, will serve as an important component of assembly formation.  Before examining how external input correlations can train the network into different macroscopic structures, we first must examine the role of inhibition and inhibitory plasticity in this network.

\subsubsection{Inhibition and homeostatic inhibitory STDP maintain stable activity}
\begin{figure}[ht!]
\centering \includegraphics{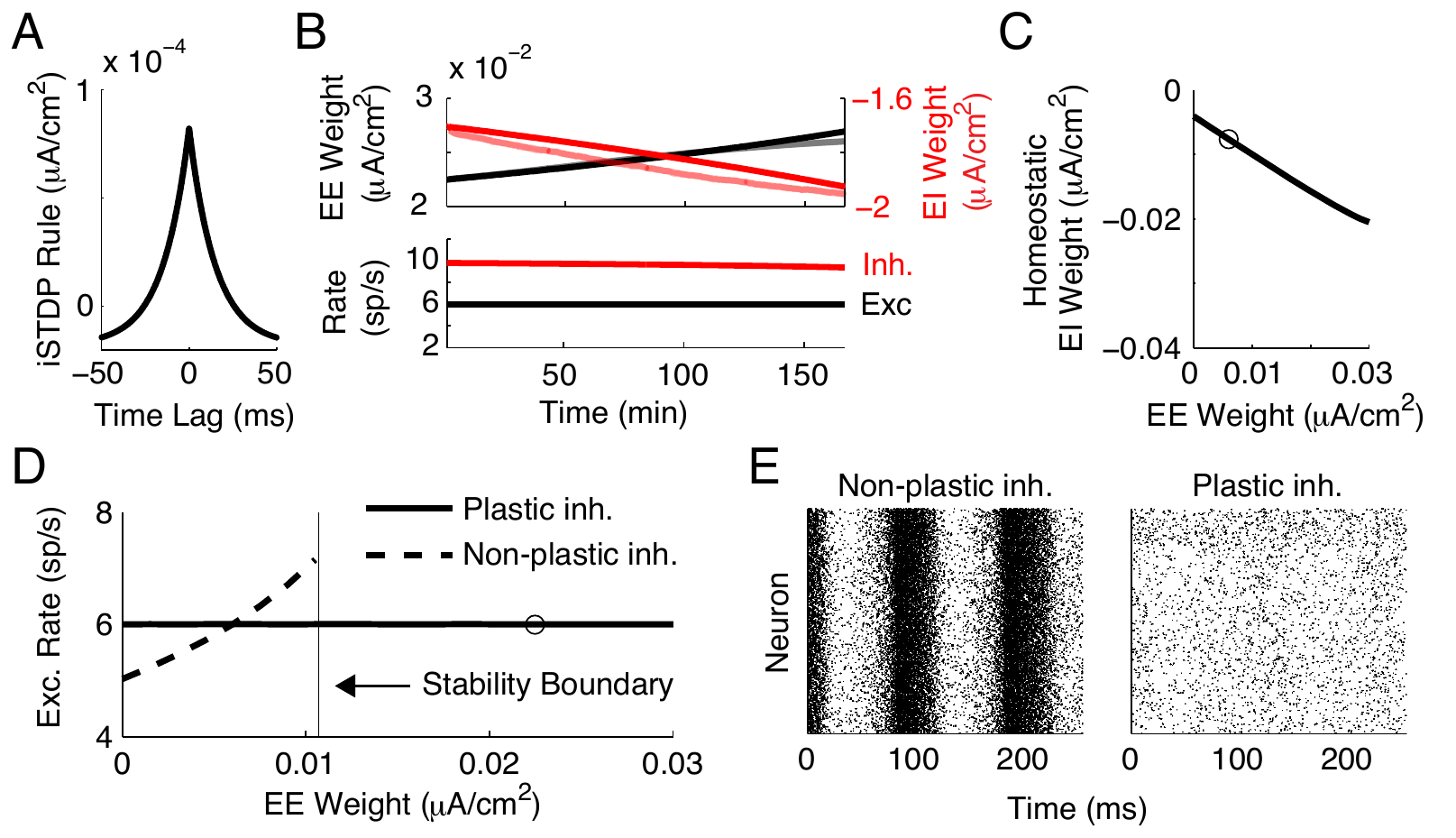}
\caption[{\bf \small Homeostatic inhibitory plasticity dynamically stabilizes firing rates.}]{
\footnotesize
{\bf Homeostatic inhibitory plasticity dynamically stabilizes firing rates.} (A) STDP rule for inhibitory-excitatory synapses.  (B)  Top: Coevolution of mean excitatory-excitatory weight $p$ (black) and mean inhibitory-excitatory $p_{EI}$ (red). Bottom: Firing rates during plasticity.  (C) The fixed point for $p_{EI}$ as a function of the mean excitatory strength $p$.  Open circle marks the inhibitory weight used for the nonplastic inhibition in later panels. (D) Firing rates as a function of excitatory weight in the cases of plastic and nonplastic inhibition.  We predicted the location of that stability boundary by numerically computing the eigenvalues of the Fokker-Planck equation associated with the single-neuron voltage distribution and examining how activity is recurrently filtered through the network \cite{ledoux_dynamics_2011}.  (E)  Raster plots of the network activity.  In both bases the excitatory weight is at the value marked by the circle in panel D.  For the right raster, $p_{EI}$ is at its homeostatic fixed point.}  
\label{Fig_assembly2}
\end{figure}

\paragraph{}
In recurrent networks, excitatory plasticity can lead to the destabilization of asynchronous activity \cite{lubenov_decoupling_2008} and the development of pathological synchrony \cite{morrison_spike-timing-dependent_2007}.  Past modeling studies have explored plasticity of inhibition as a stabilizing mechanism \cite{vogels_inhibitory_2013}, preventing runaway activity in networks with \cite{litwin-kumar_formation_2014, zenke_diverse_2015} and without \cite{vogels_inhibitory_2011} excitatory plasticity.  Recent experiments in humans using a combination of transcranial direct current stimulation and ultra-high field MRI has given evidence for an association-dependent balancing of excitation and inhibition \cite{barron2016unmasking}, where inhibitory plasticity was a suggested mechanism. Indeed, plasticity of inhibitory-excitatory connectivity maintains a balance between excitation and inhibition in layer 5 of mouse auditory cortex in vitro \cite{damour_inhibitory_2015}.  We followed these studies and, to prevent runaway excitation due to potentiation of excitatory synapses, we modeled inhibitory $\rightarrow$ excitatory homeostatic spike timing--dependent plasticity (iSTDP): pairs of near coincident pre- and postsynaptic spikes caused potentiation of inhibitory-excitatory synapses, while individual presynaptic spikes caused depression \cite{vogels_inhibitory_2011} (Figure \ref{Fig_assembly2}A).  The strength of this depression was determined by the homeostatic target excitatory rate, $\bar{r}_E$ (Methods, \ref{sec:plast}). 

\paragraph{} 
%Inhibitory feedback recruited within the network reduced the spiking activity of excitatory neurons.  With fixed inhibitory-excitatory projection strengths, potentiation of excitatory synapses could lead to runaway activity if the recurrent excitation became strong enough to overcome the inhibitory feedback \cite{morrison_spike-timing-dependent_2007, litwin-kumar_formation_2014}.  In order to prevent this, we let inhibitory-excitatory synapses obey an inhibitory STDP (iSTDP) rule \cite{vogels_inhibitory_2011, vogels_inhibitory_2013, damour_inhibitory_2015}:
%\begin{equation}
%\epsilon L_I(s)=\mathcal{H}(\mathbf{W}_{ij}-W^\text{max,I}) f_I e^{-\frac{\left| s\right|}{\tau_I}}
%\end{equation}
%where $f_I$ sets the maximum amplitude of individual potentiations and $\tau_I$ determines their dependence on relative spike timing (parameter values in Table \ref{tab:par2}). In addition, presynaptic (inhibitory) spikes cause depression by $\mathcal{H}(-\mathbf{W}_{ij}) d_I$, with $d_I = -2f_I \bar{r}_E\tau_I$. 

We took the excitatory eSTDP rule to be balanced between potentiation and depression, but if the excitatory firing rates were far from the target rate $\bar{r}_E$, then the inhibitory plasticity became unbalanced and its leading-order dynamics did not depend on internally generated spike time correlations (Methods, \ref{sec:pEIunbalstab}):
\begin{equation} \label{dpEIdt_unbal}
\frac{dp_{EI}}{dt} = \left(r_I \big(r_E-\bar{r}_E\big)S^I +c_{EI}\sigma^2 S^{EI}_\eta \right)p_0^{EI} 
\end{equation}
Together with the dynamics of the firing rates $r_E,r_I$, these occurred on a faster timescale than the balanced plasticity of excitatory connectivity.  Examining the fixed points and stability of $(p_{EI},r_E,r_I)$ on this unbalanced timescale revealed that the inhibitory plasticity stabilizes the firing rates so that $r_E - \bar{r}_E \sim \mathcal{O}(\epsilon)$ (Methods, \ref{sec:pEIunbalstab}).  Indeed, in simulations we saw that as $p$ increased (decreased), $p_{EI}$ potentiated (depressed) and maintained $r_E = \bar{r}_E + \mathcal{O}(\epsilon)$ (e.g., Figure \ref{Fig_assembly2}B).

%%%%%%%%%%%%%%%%%%%%%%%%%%%%%%%%%%%%%%%%%%%%%%%%%

%%%%%%%%%%%%%%%%%%%%%%%%%%%%%%%%%%%%%%%%%%%%%%%%%

\paragraph{}
The location of the homeostatic inhibitory weight, $p^*_{EI}$, is given by solving the leading-order dynamics of the unbalanced inhibitory plasticity for $dp_{EI}/dt = 0, dr_E/dt = 0, dr_I/dt = 0$.  Due to the separation of timescales between the homeostatic iSTDP and the balanced eSTDP, we could predict the location of the homeostatic inhibitory weight $p^*_{EI}$ through a quasi-static approximation of $p$ (Methods, \ref{sec:pEIunbalstab}).  We tracked the location of the homeostatic inhibitory weight $p_{EI}^*$ as a function of $p$.  As expected, strong recurrent excitation required stronger inhibitory-excitatory feedback to enforce  $r_E = \bar{r}_E + \mathcal{O}(\epsilon)$ (Figure \ref{Fig_assembly2}C).  In order to investigate the conditions under which inhibition was able to maintain stable activity at that homeostatic fixed point, we compared the cases of plastic and nonplastic inhibition.  With nonplastic inhibition, firing rates increased with $p$.  If the excitatory feedback $p$ became strong enough, the stationary firing rates lost stability (Figure \ref{Fig_assembly2}D).  
This instability was reflected in the development of hypersynchronous spiking (Figure \ref{Fig_assembly2}E), in contrast to the weakly correlated spiking activity in the network with plastic inhibition.  
In total, in order to study the robust potentiation of recurrent excitation, we required a counterbalancing potentiation of inhibitory onto excitatory neurons so as to homeostatically maintain a weakly correlated yet strongly connected excitatory network.  

\subsection{Stimulus-induced correlations drive assembly formation}
\paragraph{}
The thresholds for potentiation and depression in both $p$ and $q$ suggested a mechanism for the formation of assembly structure through spike timing.  Namely, if we define $p$ and $q$ variables for within- and cross-assembly connectivity, each should obey similar dynamics to Eqs. \eqref{eq:dpEEdt}, \eqref{eq:dqEEdt}.  In particular, each should have a threshold for potentiation. Furthermore, these thresholds should depend on the spatial correlation of the external inputs to within- or cross-cluster pairs of neurons.  

\begin{figure}[ht!]
\centering \includegraphics{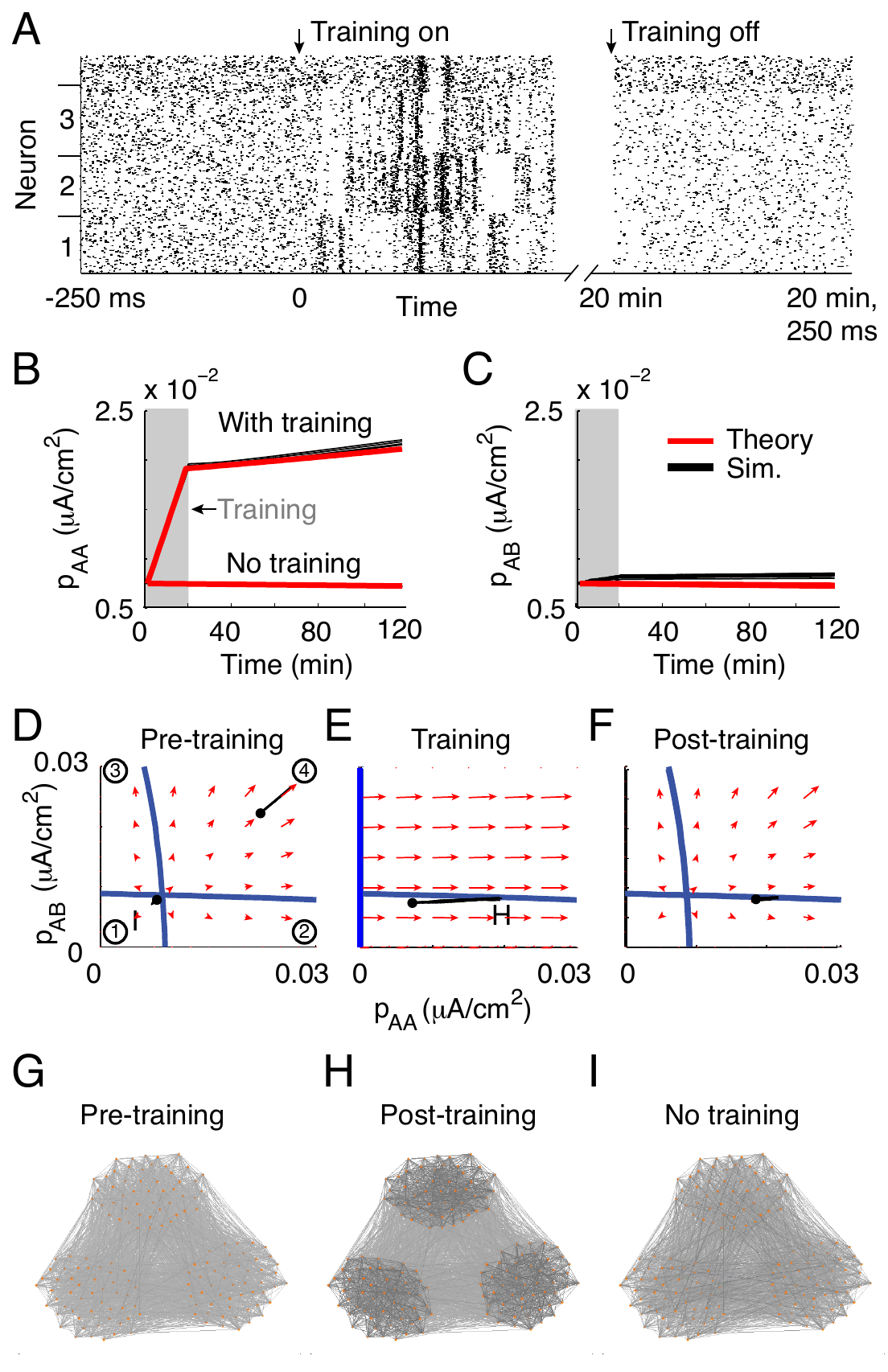}
\caption[{\bf \small Training and spontaneous reinforcement of assembly structure.}]{
\footnotesize
{\bf Training and spontaneous reinforcement of assembly structure.} (A) Spike train raster of the network activity.  Excitatory neurons are ordered by assembly membership (numbers on the ordinate axis). (B,C) Dynamics of the mean strength of within-assembly synapses (B) and of cross-assembly synapses (C).  The red lines are from our theory in Eqs. \eqref{eq:dpAAdt} and \eqref{eq:dpABdt}.  The black lines are computed from trial averaging the outputs spiking simulations.  The grey boxes mark the stimulus duration. (D--F) Phase planes governing the mean synaptic weights before the training signal (D), during the training period (E) and after training (F).  The numbers in D are referred to in the main text.  The blue curves are nullclines computed from Eqs. \eqref{dpEIdt_unbal}, \eqref{eq:dpAAdt} and \eqref{eq:dpABdt}.  (G--I) Visualization of a subset of the excitatory-excitatory connectivity.  Nodes positioned by the Fruchterman-Reingold force algorithm.}  
\label{Fig_assembly3}
\end{figure}

\paragraph{}
We began by studying the simpler case of networks with asymmetric baseline connectivity ($\Omega = 0$) so that $q$ could be neglected.  We divided the excitatory neurons into $M$ putative assemblies of $\kappa$ neurons each, based on their assigned stimulus preferences.  Each assembly contained neurons that received spatially correlated inputs due to an external stimulus (Figure \ref{Fig_assembly3}A). For ease of calculation, we assumed that the assemblies were symmetric so that the connectivity within and between assemblies was characterized by:
\begin{equation} \begin{aligned} \label{pdef}
\epsilon p_{AA} &= \frac{1}{\kappa^2}\sum_{i,j \in A} \mathbf{W}_{ij} \\
\epsilon p_{AB} &= \frac{1}{\kappa(N_E-\kappa)}\sum_{i \in A} \sum_{j \not\in A} \mathbf{W}_{ij}
\end{aligned} \end{equation}
where $p_{AA}$ is the mean strength of connections within an assembly, and $p_{AB}$ is the mean strength of all cross-assembly connections.  The correlation of the external inputs to neurons with the same (different) input preferences was $c_{AA}$ ($c_{AB}$).  The inhibitory-excitatory, excitatory-inhibitory and inhibitory-inhibitory connectivities remained unstructured and asymmetric.  Following the same steps as for $p$, we derived dynamical equations for the mean within- and cross-cluster connectivity (Methods, \ref{sec:dynamicsrbar}):
\begin{equation} \label{eq:dpAAdt}
\frac{dp_{AA}}{dt} = \left( r_E^2 S + c_{AA}\sigma^2 S_\eta\right) p_0 + \epsilon \big[S_F  p_{AA} + S_B p_0p_{AA} + S_C  \left(p_{AA}^2 + \left(M-1\right)p_{AB}^2 \right) + S_C^I  \gamma (p_{EI}^*)^2 \big]
\end{equation}
\begin{equation} \label{eq:dpABdt}
\frac{dp_{AB}}{dt} = \left( r_E^2 S + c_{AB}\sigma^2 S_\eta\right) p_0 + \epsilon \big[ S_F p_{AB} + S_Bp_0p_{AB} + S_C \left(2p_{AA}p_{AB} + \left(M-2 \right)p_{AB}^2 \right) + S_C^I  \gamma (p_{EI}^*)^2 \big]
 \end{equation}
Due to our approximation of the spike train covariances (Eq. \eqref{Ctrunc_inh}), the dynamics of the mean synaptic weight within and across assemblies are coupled to each other only through correlations due to common inputs (the $S_C$ terms).

%%%%%%%%%%%%%%%%%%%%%%%%%%%%%%%%%%%%%%%%%%%%%%%%%%

%%%%%%%%%%%%%%%%%%%%%%%%%%%%%%%%%%%%%%%%%%%%%%%%%%

\paragraph{}
Numerical solution of Eqs. \eqref{eq:dpAAdt} and \eqref{eq:dpABdt} showed that training with $c_{AA}>0$ and $c_{AB}=0$ (20 minutes) produced robust assembly formation ($p_{AA}$ increased, Figure \ref{Fig_assembly3}B red lines) while maintaining low cross-assembly coupling ($p_{AB}$ remained low, Figure \ref{Fig_assembly3}C red lines).  Furthermore, after training the assembly structure self-reinforced, with $p_{AA}$ continuing to increase even though $c_{AA}=0$ (Figure \ref{Fig_assembly3}B,C outside the grey shaded area).  These results are a main contribution of our study and represent a spike timing--based mechanism for assembly formation (Figure \ref{Fig_schematic}C) as an alternative to rate-based mechanisms (Figure \ref{Fig_schematic}B).  We next used the low dimensionality of Eqs. \eqref{eq:dpAAdt} and \eqref{eq:dpABdt} to analyze the dynamical mechanisms by which assembly formation occurred.        

\paragraph{}
Similar to the network without assembly structure, the nullclines of $p_{AA}$ and $p_{AB}$ predicted their thresholds for potentiation.  
%The form of the dynamics suggests that the $p_{AA}$ and $p_{AB}$ nullclines could be simply computed by solving those quadratic equations.  This procedure is complicated, however, by the fact that the homeostatic inhibitory weight depends on both $p_{AA}$ and $p_{AB}$.  
In order to numerically compute (for example) the $p_{AA}$ nullcline, we found for each $p_{AA}$ the $p_{AB}$ that, in combination with the induced inhibitory weight $p_{EI}^*$, yielded $dp_{AA}/dt = 0$.  Assuming that the eSTDP rule is temporally symmetric ($\tau_- \sim \tau_+ + \mathcal{O}(\epsilon)$) so that $S_C$ and $S_C^I$ both vanish permits an explicit calculation of the nullclines (Methods, \ref{sec:tempsym}):
\begin{equation}
p_\alpha^* = -\frac{\left(\bar{r}_E^2 S + c_\alpha\sigma^2S_\eta\right)p_0}{\epsilon \left(S_F+p_0S_B\right)} \label{eq:pstar}
\end{equation} 
for $\alpha \in \{AA,AB\}$.  This then gives horizontal and vertical nullclines in ($p_{AA}, p_{AB}$) space, effectively decoupling the $p_{AA}$ and $p_{AB}$ dynamics.  While this assumption is quantitatively inaccurate for our STDP rule (which has $\tau_- = 2\tau_+$), it reveals the main effect of external input correlations.  Note that while the small parameter $\epsilon$ appears in the denominator, both terms of the numerator are also $\mathcal{O}(\epsilon)$ due to the balance between potentiation and depression in the eSTDP rule (i.e $S,S_\eta \sim \mathcal{O}(\epsilon)$).  

\paragraph{}

Eq. \eqref{eq:pstar} shows that $p_\alpha^*$ is positive in the absence of external input correlations ($c_\alpha = 0$).  This is because we took $S < 0$ and the network is not fully connected ($p_0 < 1$), so that $S_F +p_0S_B > 0$.  In the absence of training, the fixed point $(p_{AA}^*,p_{AB}^*)$ was unstable and the nullcline structure partitioned ($p_{AA}$, $p_{AB}$) space into four quadrants (Figure \ref{Fig_assembly3}D): 1) a region where all structure dissolved because $p_{AA}$ and $p_{AB}$ both depressed, 2) a region where assembly structure formed since $p_{AA}$ potentiated while $p_{AB}$ depressed, 3) a region where a loop between assemblies formed because $p_{AB}$ potentiated while $p_{AA}$ depressed, and finally 4) a region where assemblies fused since $p_{AA}$ and $p_{AB}$ both potentiated.  With $c_\alpha > 0$, the nullcline $p_\alpha^*$ was decreased by an amount proportional to $c_\alpha$. In particular, $c_{AA} > 0$ reduced the threshold for potentiation of within-assembly connectivity, while leaving the threshold for cross-assembly connectivity unaffected.  Thus, when an initial state was in region 1, training with $c_{AA} > 0$ and $c_{AB} = 0$ would result in the dynamics shifting to region 2 so that assembly structure formed (Figure \ref{Fig_assembly3}D-F).  Once training was completed, if $p_{AA}$ increased sufficiently, then the state post-training remained in quadrant 2 and assembly structure continued to form, albeit at a slower rate.  Thus an analysis of the mean field theory of Eqs. \eqref{eq:dpAAdt} and \eqref{eq:dpABdt} gives a qualitative understanding of the dynamics of assembly formation.   

\paragraph{}
We tested these mean field theory predictions in simulations of the full system of spiking neurons, divided into $M=3$ assemblies.  After 20 min of stimulation, we observed the formation of strongly connected assemblies of neurons (Figure \ref{Fig_assembly3}G,H).  The connectivity between assemblies was not potentiated; the assemblies did not fuse.  We contrast this to the same network after 20 min of spontaneous activity: structure did not form spontaneously (Figure \ref{Fig_assembly3}I).  Furthermore, the mean field theory of Eqs. \eqref{eq:dpAAdt} and \eqref{eq:dpABdt} gave an excellent match to the $p_{AA}$ and $p_{AB}$ estimated from the spiking network simulations (Figure \ref{Fig_assembly3}B,C, black versus red curves).  In total, our low-dimensional mean field theory not only gives a qualitative understanding of assembly formation through spike timing, but also gives a quantitatively accurate theory for the high-dimensional spiking network simulations upon which the theory is based.   

\paragraph{}

\begin{figure}[ht!]
\centering \includegraphics{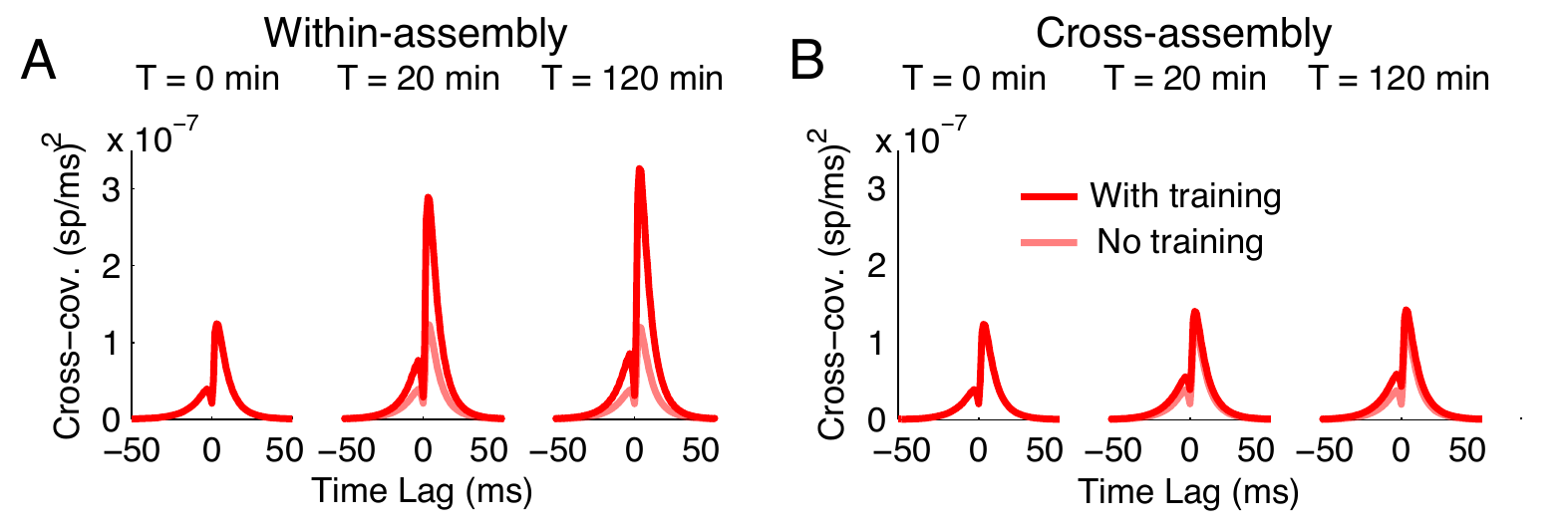}
\centering \caption[{\bf \small Spike train covariability reflects and reinforces learned network structure.}]{
\footnotesize
{\bf Spike train covariability reflects and reinforces learned network structure.} (A) Average spike train cross-covariance between within-assembly pairs of neurons.  (B) Average spike train cross-covariance between cross-assembly pairs of neurons.  Cross-covariances estimated by the truncated linear response theory, Eq. \eqref{Ctrunc_inh}.  Solid: with training.  Shaded: without training.  Left: before training.  Middle: end of stimulus presentation.  Right: after spontaneous activity following stimulus presentation (as in Figure \ref{Fig_assembly3}).  
} 
\label{Fig_assembly4}
\end{figure}

Finally, while the synaptic strengths $p_{AA}$ and $p_{AB}$ evolved on a slow timescale of minutes, the STDP rule is sensitive to spike time correlations on a fast timescale of tens of milliseconds.  Internally generated spike time correlations depend upon the recurrent network structure, and hence the covariance between neuron spike trains reflected the slow changes in $p_{AA}$ and $p_{AB}$. Indeed, spiking covariability after training was much larger within assemblies than between them (Figure \ref{Fig_assembly4}A,B).  Further, these differences were reinforced post-training, reflecting the concomitant dynamics of $p_{AA}$ and $p_{AB}$ during this time.  Thus, the malleability of internal correlations provided a signature of assembly formation observable in the fast-timescale dynamics of coordinated spiking activity.

%\subsection{Spontaneous spiking covariability after learning reinforces learned structure}
%\paragraph{}
%In the absence of a training signal, assemblies did not spontaneously form.  Our theory predicted, however, that a threshold for assembly formation did exist under spontaneous spiking conditions (Figure \ref{Fig_assembly3}D).  This suggested that if the assemblies potentiated sufficiently during the stimulus presentation to pass the location of that threshold during spontaneous activity, then internally generated spiking covariances would reinforce the learned architecture after stimulus removal.  internally generated spiking covariability after training was much larger within assemblies than between them (Figure \ref{Fig_assembly4}).  We predicted that this asymmetry in spiking covariability, due to the learned network structure, would reinforce that structure after the end of training.
%
%\paragraph{}
%We tested this prediction by examining the evolution of $(p_{AA},p_{AB})$ for 100 min after stimulus removal in our spiking simulations.  The learned assembly structure was stable, and further reinforced by internally generated spiking covariability after the removal of the training signal (Figure \ref{Fig_assembly3}B,D,F).  This reinforcement was also reflected by the further increase of spiking correlations between within-assembly neuron pairs (Figure \ref{Fig_assembly4}).  Internally generated correlations, because they reflected the trained structure of the network, further enforced that architecture.

%%%%%%%%%%%%%%%%%%%%%%%%%%%%%%%%%%%%%%%%%%%%%%%%%
%%%%%%%%%%%%%%%%%%%%%%%%%%%%%%%%%%%%%%%%%%%%%%%%%

\subsection{Reciprocal excitatory connectivity is preferentially promoted between similarly tuned neurons}
\paragraph{}
In the previous section we examined how spatial correlations in external signals can promote the formation of neuronal assemblies.  We discussed this only at the level of mean synaptic weights, the simplest measure of connectivity between neuron pairs.  Recent data have revealed another striking feature of pair-based connectivity: pairs of neurons with similar stimulus preferences have strong reciprocal connectivity \cite{cossell_functional_2015}.  Theories of STDP focusing on pairs of neurons suggest that Hebbian STDP should suppress reciprocal connectivity \cite{song_competitive_2000, babadi_pairwise_2013} (but see \cite{tannenbaum2016shaping}). Our previous work has suggested that if reciprocal connectivity is sufficiently strong in a network on average, it can be reinforced by Hebbian STDP \cite{ocker_self-organization_2015}. We next examined whether plasticity driven by precisely correlated spike times could contribute to the development of strong reciprocal connectivity in neuronal assemblies.

\paragraph{}
To that end we considered networks with partially symmetric baseline connectivity ($\Omega = 0.4$).  This reciprocal structure is reflected in the weight matrix $\mathbf{W}$.  To measure it in a way that allows us to take into account the development of stimulus-driven assemblies, we consider two metrics of the network structure in addition to $p_{AA}$ and $p_{AB}$:
\begin{equation} \begin{aligned}
\epsilon q_{AA} &= \frac{1}{\kappa^2} \sum_{i,j \in A} \mathbf{W}^0_{ij} \mathbf{W}_{ji} - p_0 \epsilon p_{AA} \\
\epsilon q_{AB} &= \frac{1}{\kappa\left(N_E-\kappa\right)} \sum_{i \in A} \sum_{j \not\in A} \mathbf{W}^0_{ij} \mathbf{W}_{ji} - p_0 \epsilon p_{AB}
\end{aligned} \end{equation}
These measure the average strength of reciprocal connections either within ($q_{AA}$) or between ($q_{AB}$) assemblies, above what would be expected by chance.  As before, we assume symmetry between different assemblies.  The inclusion of the mean reciprocal weights expands our description of the network structure to four dimensions $(p_{AA},p_{AB},q_{AA},q_{AB})$.  Furthermore, the dynamics of the mean synaptic weights $p_{AA}$ and $p_{AB}$, in addition to depending on each other, now depend on $q_{AA},q_{AB}$ through the STDP-weighted covariances due to reciprocal connections (see Methods \ref{sec:dynamicsrbar}).  

%%%%%%%%%%%%%%%%%%%%%%%%%%%%%%%%%%%%%%%%%%%%%%%%%
\begin{figure}[ht!] 
\centering \includegraphics{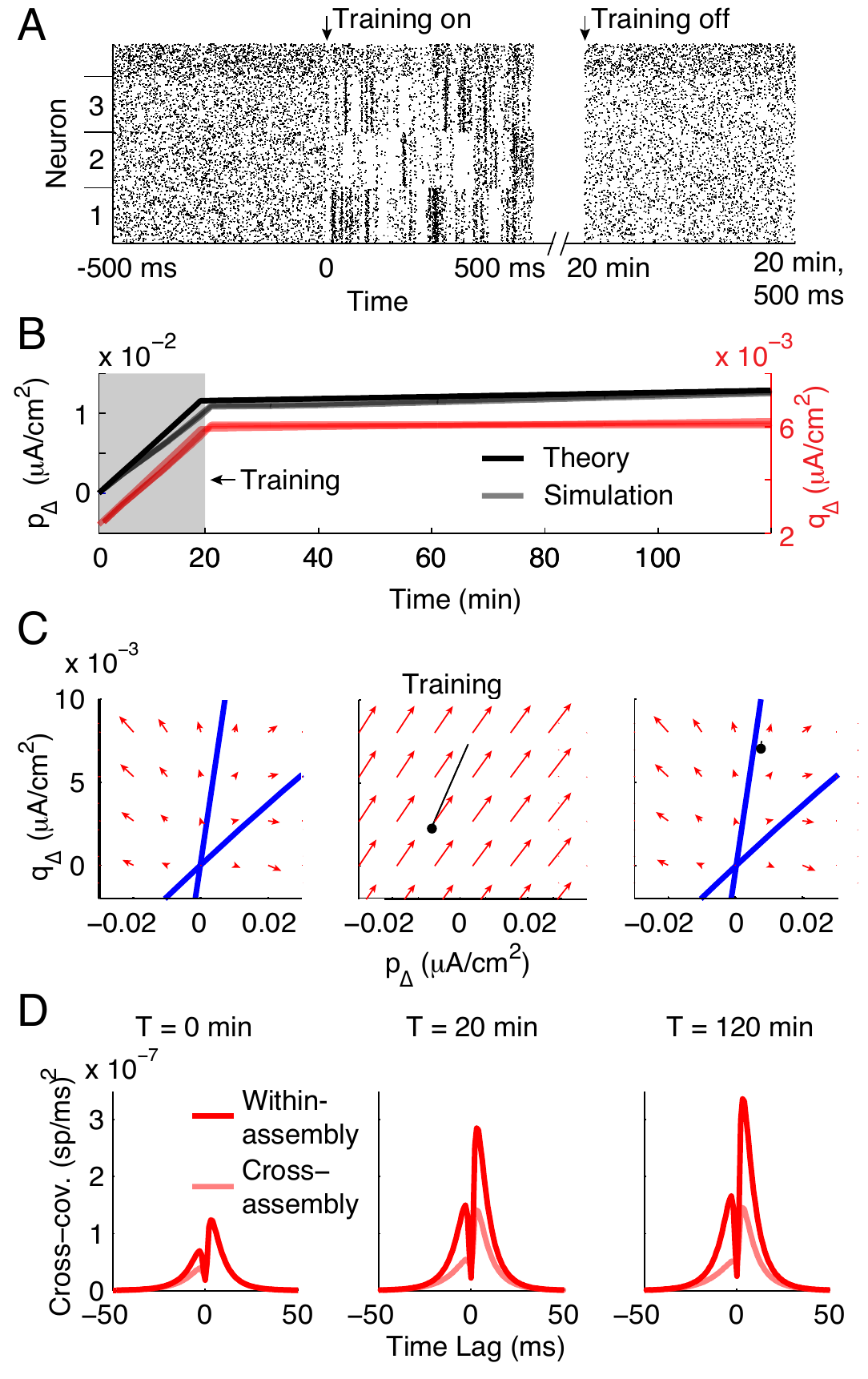}
\caption[{\bf \small Reciprocal connectivity is preferentially promoted within assemblies.}]{
\footnotesize
{\bf Reciprocal connectivity is preferentially promoted within assemblies.} We consider networks with partially symmetric baseline connectivity.  (A) Raster of network activity during pre-training, training and post-training phases.  Excitatory neurons ordered by assembly membership (labeled on ordinate axis).  (B)  Time course of the relative strength of within-assembly synapses, $p_\Delta$ (black), and within-assembly reciprocal synapses, $q_\Delta$ (red).  (C)  Phase plane of $p_\Delta,q_\Delta$ in the absence (pre- and post-training) or presence (training) of external input correlations.  Blue: nullclines for $c_\Delta = 0$ were $p_\Delta^* = \frac{-(S_F+p_0S_B) \pm \sqrt{\left(S_F+p_0S_B\right)^2 - 4S_CS_Bq_\Delta}}{2 S_C}$ and $q_\Delta^* = -\frac{S_B(1-p_0)p_0p_\Delta + S_C\frac{q_0}{p_0}p_\Delta^2 }{S_F+S_B(1-p_0)}$.  Black traces: simulation of the spiking network.   (D) Average spike train cross-covariances (truncated approximation, Eq. \eqref{Ctrunc_inh}).  Left, before training.  Middle, immediately at end of training.  Right, 100 min post-training.
} 
\label{Fig_assembly5}
\end{figure}
%%%%%%%%%%%%%%%%%%%%%%%%%%%%%%%%%%%%%%%%%%%%%%%%%

\paragraph{}
In order to obtain a simpler description, we considered the change of variables:
\begin{equation} \begin{aligned}
p_\Delta &= p_{AA} - p_{AB} \\
q_\Delta &= q_{AA} - q_{AB}.
\end{aligned} \end{equation}
These measure the relative strength of assembly structure in the network, at the levels of mean connection strength ($p_\Delta$) and above-chance reciprocal connection strength ($q_\Delta$).  In order for a network to respect the structure observed in mouse V1 by Cossell et al. \cite{cossell_functional_2015}, it should have $p_\Delta > 0, q_\Delta > 0$.  The dynamics of $(p_\Delta,q_\Delta)$ can be simply calculated from those of $(p_{AA},p_{AB},q_{AA},q_{AB})$ and are (see Methods \ref{sec:Deltapq}):
\begin{equation}
\frac{dp_\Delta}{dt} = c_\Delta \sigma^2S_\eta p_0 + \epsilon \left[S_Fp_\Delta + S_B(q_\Delta + p_0p_\Delta) + S_Cp_\Delta^2 \right] \label{eq:deltap}
\end{equation}
\begin{equation}
\frac{dq_\Delta}{dt} = c_\Delta \sigma^2S_\eta q_0 + \epsilon \left[S_Fq_\Delta + S_B\left(1-p_0\right)\left(q_\Delta + p_0p_\Delta \right) + S_C\frac{q_0}{p_0}p_\Delta^2  \right] \label{eq:deltaq}
\end{equation}
where $c_\Delta = c_{AA} - c_{AB}$.  Notably, the dynamics of $p_\Delta$ and $q_\Delta$ decoupled from the overall strengths of excitation and inhibition in the network, i.e., Eqs. \eqref{eq:deltap} and \eqref{eq:deltaq} do not explictly depend on $p_{AA}, p_{AB}, q_{AA}$ and $q_{AB}$.  
%The homeostatic inhibitory STDP, for example, is still in effect and maintains excitatory firing rates at $\bar{r}_E$.  Common inputs from inhibitory neurons, however, contribute equally to the spike train covariance of within- and cross-assembly excitatory neuron pairs.  Because of that symmetry, they do not affect the differential plasticity of assembly structure, $p_\Delta$ or $q_\Delta$.  
Further, the contribution of chance spike coincidences, $r_E^2 S$, canceled because neurons in each assembly have the same average firing rate.  Satisfyingly, the mean field theory of Eqs. \eqref{eq:deltap} and \eqref{eq:deltaq} gave an accurate match to network simulations during training ($c_{\Delta} >0$) and spontaneous ($c_{\Delta} =0$) regimes (Figure \ref{Fig_assembly5}A,B).    

\paragraph{}
Similar to the case of asymmetric networks, these dynamics admit nullclines that represent thresholds for potentiation/depression (Figure \ref{Fig_assembly5}C, blue curves).  %With $c_\Delta = 0$, these are given by:
%\begin{equation} \begin{aligned}
%p_\Delta^* &= \frac{-(S_F+p_0S_B) \pm \sqrt{\left(S_F+p_0S_B\right)^2 - 4S_CS_Bq_\Delta}}{2 S_C} \\
%q_\Delta^* &= -\frac{S_B(1-p_0)p_0p_\Delta + S_C\frac{q_0}{p_0}p_\Delta^2 }{S_F+S_B(1-p_0)}
%\end{aligned} \end{equation}
The origin $(p_\Delta=0,q_\Delta=0)$ is unstable  %(the other equilibrium point is also unstable, but is at inaccessible values of $(p_\Delta,q_\Delta)$).  
and the nullclines divide the phase plane into four regions, containing each potential combination of potentiation and depression of $(p_\Delta, q_\Delta)$.  We take the synaptic weights to be initially unstructured, so that before training $p_\Delta \approx q_\Delta \approx 0$ (Figure \ref{Fig_assembly5}C, left).  If external input correlations are higher for within-assembly pairs than cross-assembly pairs ($c_\Delta > 0$), the unstable point at $(0,0)$ is shifted to negative $(p_\Delta,q_\Delta)$ (Figure \ref{Fig_assembly5}C, middle).  This pushed the unstable synaptic dynamics towards having assemblies of strongly reciprocally connected neurons (Figure \ref{Fig_assembly5}C, right). %This occurred both for the relative strength of reciprocal connections $q_\Delta$ and for the mean strength of within-assembly connections, $q_{AA}$.

\paragraph{}
This shift in network structure was reflected by the magnitude of spike train covariances within and between assemblies.  Indeed, the training of assembly structure into the network led to a doubling of spike train covariability for within-assembly neurons compared to cross-assembly neurons (Figure \ref{Fig_assembly5}D).  Due to the higher levels of reciprocal connectivity, the average spike train covariances at negative time lags were larger than for the network with $\Omega = 0$ (compare Figure \ref{Fig_assembly5}D vs Figure \ref{Fig_assembly4}).  As was the case for asymmetric networks, these results suggest that spontaneously generated spike train correlations, in addition to providing a signature of learned network structure, can actively reinforce it.  

\subsection{Trained noise covariance maintains coding performance}
\begin{figure}[ht!]
\centering \includegraphics{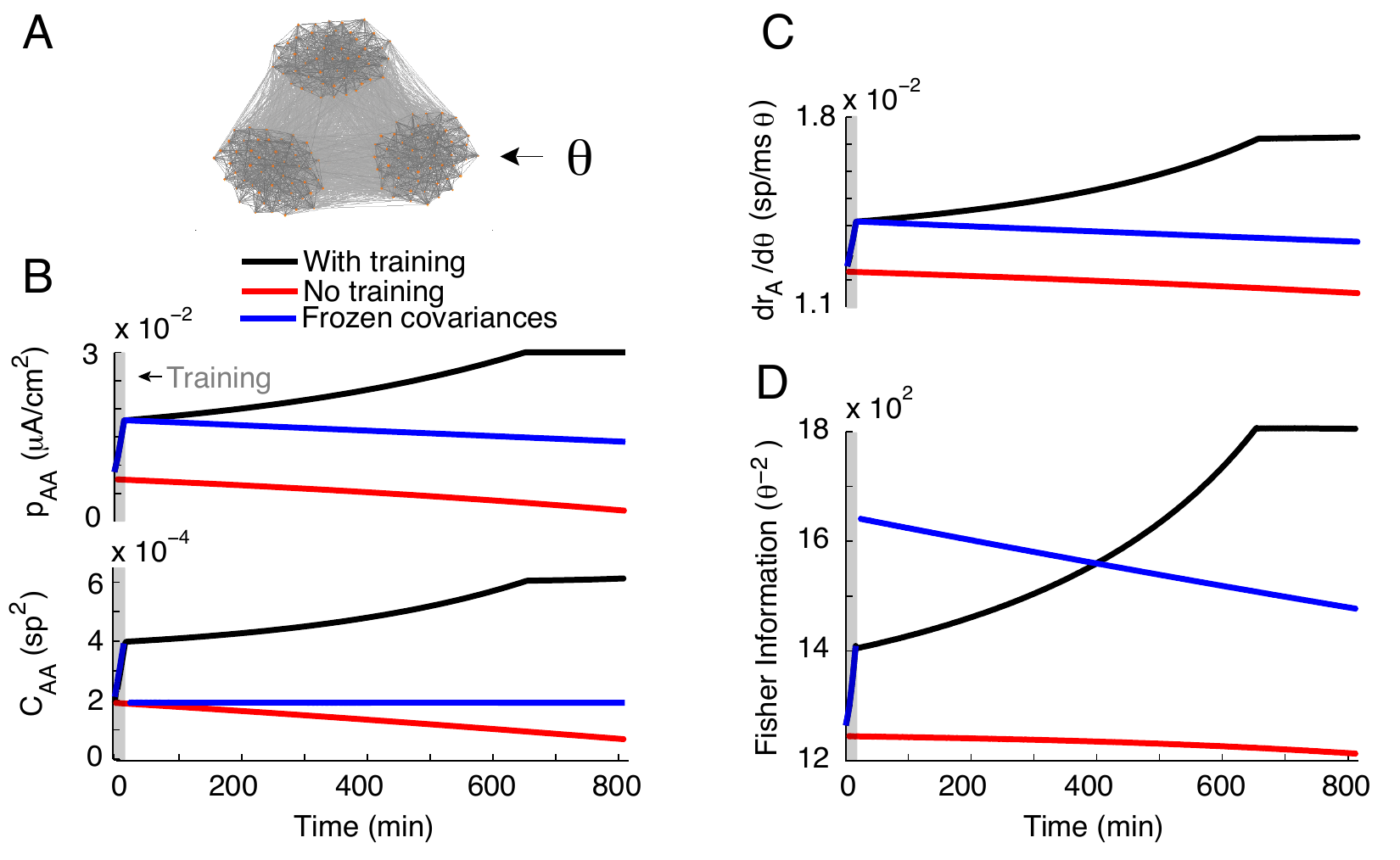}
\centering \caption[{\bf \small Training and spontaneous reinforcement of coding performance.}]{
\footnotesize
{\bf Spike train covariability reflects and reinforces learned network structure.} (A) One assembly received a stimulus, $\theta$, which it encodes by that assembly's total spike count in $T=100$ ms.  (B) Top: Variance of the summed spike count of assembly $A$ increased during and after training (black). Assembly $A$'s spike count variance decreased without training (red). As a control, we reset spiking covariability after training to pre-training values and froze them (blue). Bottom: The gain of stimulated neurons with respect to. $\theta$ increased during and after training (black). Without training, the stimulus-response gain decreased (red). With frozen covariances, the gain decreased after training (blue).  Grey box: training period.  (C) The mean strength of within-assembly connectivity. (D) Fisher information of the assembly's spike count about $\theta$.
} 
\label{Fig_assembly6}
\end{figure}

\paragraph{}
We finally asked how the spontaneous reinforcement of learned network structures, and the associated internally generated spike train covariability, affected the ability of cell assemblies to encode their preferred inputs. We took the partially symmetric network (Figure \ref{Fig_assembly5}) and allowed the external input to excitatory neurons in an assembly to depend on a stimulus $\theta$: $\mu_\mathrm{E} = \mu_\mathrm{ext} + \mu_\theta$. For simplicity, we took each stimulus to target exactly one assembly and considered only the coding by a single assembly (labeled $A$; Figure \ref{Fig_assembly6}A). 

\paragraph{}

We measured the linear Fisher information \cite{beck_linearfisher_2011} of an assembly's net activity $n_A=\sum_{i\in A} n_i$ about the stimulus $\theta$:
\begin{equation}
FI_{A} = \left(\frac{dn_A}{d\theta}\right)^2 C_{AA}^{-1}
\end{equation}
where $n^i_A$ is the spike count (over $T=100$ ms) from neuron $i$ of assembly $A$, and $C_{AA}$ is the variance of $n_A$. $dn_A/d\theta$ is the mean stimulus-response gain of the spike count of neurons in assembly $A$, so that $n_A = \kappa T r_A$ where $r_A$ is the mean firing rate of a neuron in assembly $A$. Fisher information is a lower bound on the variance of any estimate of $\theta$ from $r_A$, and the restriction to linear Fisher information gives a natural decomposition of $FI_{A}$ into the response gain in $dr_A/d\theta$ and response noise $C_{AA}$.  Since $r_A$ naively sums the assembly activity we have that for large $N$, the response variance $C_{AA} \propto \langle \textrm{Cov}(n^i_A,n^j_A) \rangle_{ij}$, meaning that the mean pairwise covariance between neurons in an assembly is the dominate contribution to the noise in $n_A$'s estimate of $\theta$.   

%%%%%%%%%%%%%%%%%%%%%%%%%%%%%%%%%%%%%%%%%%%%%%%%%
%%%%%%%%%%%%%%%%%%%%%%%%%%%%%%%%%%%%%%%%%%%%%%%%%

\paragraph{}
We compared stimulus coding in networks with and without trained network structure (Figure \ref{Fig_assembly6}B top, black versus red curves).
Since training increased spike train covariances for within-assembly pairs (Figure \ref{Fig_assembly5}D) networks, then the variance of the assembly's summed spike count also increased with training and was reinforced after training (Figure \ref{Fig_assembly6}B bottom, black curve).  In agreement, when the training signals were absent and assemblies did not form, then $C_{AA}$ slowly decreased over time (Figure \ref{Fig_assembly6}B bottom, red curve). One expectation from increased variability is that training assembly structure would, in our simplified coding scenario, be deleterious to stimulus coding (since $FI_{A} \propto C_{AA}^{-1}$).  

\paragraph{}

To determine the net impact of training on $FI_{A}$, we first evaluated how training and covariability affected response gain, to then be combined with $C_{AA}$ to ultimately yield $FI_{A}$. 
We calculated the stimulus-response gain $dn_A/d\theta$, taking into account direct stimulus-driven inputs and indirect filtering of the stimulus through recurrence onto assembly $A$ (see Methods \ref{methods: fisher}).  A consequence of trained assembly structure was increased gain through the positive feedback inherent within an assembly (Figure \ref{Fig_assembly6}C, black versus red curves).  The increased gain outweighed the increased variability so that overall $FI_{A}$ grew with training and further increased through the assembly reinforcement post-training (Figure \ref{Fig_assembly6}D, black versus red curves).  In total, assembly formation was overall beneficial to network coding despite requiring larger overall network variability.  

\paragraph{}

As a final illustration of the beneficial role of noise correlations for stimulus coding, we considered an artificial network where training occurred, yet immediately after training we reset the spike train covariability to its pre-training value and forced it to remain at this value  (Figure \ref{Fig_assembly6}B, blue curves).  For a period of time post-training the $FI_{A}$ from this network was larger than that of the trained network without a reseted covariability, owing to the combination of a large gain from training and low variability through the artificial reset (Figure \ref{Fig_assembly6}D, blue versus black curves).  However, a consequence of low spike train covariability was a slow but clear degradation of assembly structure so that response gain reduced over time.  This eventually reduced $FI_A$  so that the network with internally generated covariability showed higher $FI_A$ for times $> 400$ minutes after training.  Thus, while noise correlations can have a detrimental impact on stimulus coding, the benefits of stimulus-specific recurrent structure and the role of spike train correlations play in maintaining that structure are such that noise correlations were beneficial in our simplified coding  scenario.

\section{Discussion}
\paragraph{}
Theoretical work with eSTDP in cortical networks first established the role of timing in the formation of feedforward structures \cite{masuda_formation_2007, takahashi_self-organization_2009, gerstner_neuronal_1996, fiete_spike-time-dependent_2010,tannenbaum2016shaping}.  More recently, eSTDP has been shown to promote the spontaneous formation of structured circuit motifs \cite{ocker_self-organization_2015} as well as support the stability of attractor network structure \cite{wei2014long}.  However, the role of spike timing in the formation of trained macroscopic assembly structure has been elusive.      
We derived a low-dimensional mean field theory for the plasticity of neuronal assemblies in partially symmetric networks of integrate-and-fire neurons with excitatory STDP and homeostatic inhibitory STDP. This revealed that internally generated spike train correlations can provide a threshold for potentiation or depression of mean synaptic weight and for mean reciprocal connectivity. Spatial correlations in external inputs shifted these thresholds, promoting an assembly structure in the network.  Furthermore, the post-training structure of spike train correlations reflected the learned network structure and actively reinforced the architecture. This promoted strong synaptic weights within assemblies and strong reciprocal connectivity within assemblies.

\subsection{Rate-based versus timing--based assembly formation}
\paragraph{}
Since early seminal work \cite{bell_synaptic_1997,markram_regulation_1997,bi_synaptic_1998} there has been intense research in the role of spike timing in shaping synaptic strength \cite{markram_history_2011,markram_spike-timing-dependent_2012}.  While much theoretical work focused on phenomenological eSTDP plasticity rules (like the one used in our study) \cite{babadi_pairwise_2013,song_competitive_2000,kempter_hebbian_1999}, there have also been advances in biophysically based models of eSTDP \cite{rubin_calcium_2005,clopath_voltage_2010, graupner_calcium-based_2012,shouval_spike_2010}.  These realistic models capture the known firing rate--dependence of eSTDP \cite{sjostrom_rate_2001}, complicating the discussion surrounding the role of spike timing in synaptic learning.  Indeed, past work in recurrently coupled networks of spiking neuron models has shown that the rate-dependence of these models can be sufficient for forming and maintaining neuronal assembly structure \cite{litwin-kumar_formation_2014, zenke_diverse_2015}.  Our study gives an alternative framework, where it is the fine-timescale correlations in spiking activity that drives assembly formation.  

\paragraph{}
The mechanisms behind rate-based and timing--based assembly formation are distinct.  In the rate-based scenario, the training of assemblies is {\it sequential}---each stimulus is presented in the absence of other stimuli so that neuron pairs within the same assembly can have coordinated high firing rates to drive potentiation, while neuron pairs in different assemblies can have a high-low firing rates that drive depression.  By contrast, in the timing-based framework, assemblies can be trained in {\it parallel} since within- and cross-assembly neuron pairs can simultaneously receive correlated and uncorrelated external inputs.  Further, while both frameworks show a spontaneous reinforcement of assembly structure, the mechanics of reinforcement are quite different.  In rate-based assembly formation the learned network structure is a {\it stable} attractor in the space of synaptic weights.  Spontaneous reinforcement occurs if the network has not converged to the attractor during training \cite{litwin-kumar_formation_2014, zenke_diverse_2015}. If the network structure is perturbed from the attractor, then spontaneous activity will retrain the network \cite{litwin-kumar_formation_2014}.  In our timing-based formation, it is the position of an unstable {\it repeller} that determines the growth or decay of structure.  Spontaneous reinforcement occurs when the synaptic state is such that the repeller pushes dynamics towards more structured assembly wiring.

\paragraph{}
Unstable solutions in synaptic learning are a reflection of a competitive synaptic interaction often associated with additive Hebbian STDP rules \cite{vanRossum_stable_2000,song_competitive_2000}.  Past studies have used this instability to drive feedforward structure \cite{masuda_formation_2007, takahashi_self-organization_2009, gerstner_neuronal_1996, fiete_spike-time-dependent_2010}, in effect harnessing the causality-rewarding nature of the Hebbian rule.  Our work shows that this competitive synaptic dynamic can also be used to drive assembly structure.  At the surface this seems counterintuitive since assembly dynamics are thought of as cooperative (within the assembly). In our model this is misleading and the competition between different subcomponents of spike train covariability (forwards, backwards and common synaptic wiring) supports robust assembly wiring.  A recent study that uses a similar theoretical framework to ours has also shown how balanced STDP ($S\sim \mathcal{O}(\epsilon)$) can support the spontaneous emergence of assembly structure in small networks \cite{tannenbaum2016shaping}.  However, the STDP rule of that study was acausal, with near coincident spikes strengthening both forward and backward connections. In that study the dynamics of assembly formation did not rely on competitive synaptic interactions and there is no relationship to external stimuli; it is thus quite distinct from that exposed in our study.  

\paragraph{}
While the mechanisms underlying rate- and timing-based assembly formation via eSTDP are distinct, that is not to say that they are mutually exclusive.  Synaptic plasticity clearly has rate- and timing-based components and both can actively reinforce assembly structure, suggesting that the mechanisms may be cooperative.  Future work should investigate their interactions during learning.

\subsection{Inhibitory plasticity and inhibitory stabilization}
\paragraph{}
Inhibitory feedback plays two main roles in this study.  The first is to modulate excitatory plasticity by contributing to spike train covariability amongst excitatory neurons.  The strength of this contribution is governed by the strength of the inhibitory feedback, which is in turn governed by inhibition's second role: homeostatic control of firing rates.  Inhibition's role in stabilizing network activity in the face of strong recurrent excitation has been the focus of much recent work in theoretical neuroscience.  Notably, strong inhibitory feedback provides dynamical explanations for the generation of variable and asynchronous activity \cite{van_vreeswijk_chaotic_1998, renart_asynchronous_2010, tetzlaff_decorrelation_2012, helias_correlation_2014} and can also account for paradoxical responses to external inhibitory inputs \cite{tsodyks_paradoxical_1997} and diverse features of tuning in visual cortex \cite{rubin_stabilized_2015}.

\paragraph{}
In the absence of inhibition, potentiation of excitatory synapses in our networks led to runaway excitation, meaning that in the presence of inhibition the network existed in an inhibitory-stabilized regime (Figure \ref{Fig_assembly2}).  In contrast to other recent studies \cite{litwin-kumar_formation_2014, zenke_diverse_2015}, inhibitory STDP alone was sufficient to stabilize the network activity in our work without imposing synaptic scaling or other compensatory mechanisms.  This was due to the relationship between the timescales of excitatory and inhibitory plasticity.  We take the excitatory plasticity to be balanced between potentiation and depression.  This sets the dynamics of the mean excitatory weight $p$ to occur on an $\mathcal{O}(\epsilon^{-1})$ timescale set by the eSTDP rule and the magnitude of spike train correlations.  When the firing rates are maintained at their stable fixed points, the inhibitory STDP is similarly governed by a timescale set by the iSTDP rule and the magnitude of spike train correlations.  If the firing rates are outside an $\mathcal{O}(\epsilon)$ neigborhood around their  fixed point, this causes the iSTDP rule to become unbalanced, so that it is governed by an $\mathcal{O}(1)$ timescale (Methods, \ref{sec:pEIunbalstab}).  This feature---that the inhibitory STDP can become unbalanced in order to maintain stable activity---guarantees that it can dynamically stabilize the network activity in the face of the balanced excitatory plasticity.

\paragraph{}
The question of how neurons can undergo associative, Hebbian learning while maintaining stable activity has long been studied \cite{miller_role_1994, miller_synaptic_1996}.  While homeostasis is often thought of as a slower process than learning, recent work has highlighted the necessity of homeostatic mechanisms operating on a comparable timescale to excitatory plasticity \cite{zenke_synaptic_2013}.  Homeostatic regulation acting alone, however, can paradoxically destabilize network activity, inducing oscillations in neurons' firing rates \cite{harnack_stability_2015}.  Homeostatic regulation mediated by diffusive transmitters like nitrous oxide can have different effects than that mediated by synaptic mechanisms \cite{sweeney_diffusive_2015}. The study of how homeostatic regulation and mechanisms for associative learning interact to allow stable memories and stable activity remains an exciting area of open inquiry.

\subsection{Correlated spontaneous activity can maintain coding performance}
\paragraph{}

Many theoretical studies have asked how the joint trial-to-trial fluctuations in population response (noise correlations) impact population coding \cite{zohary_correlated_1994, averbeck_neural_2006, kohn2016correlations}.  The answer to this question depends on many factors.  In particular, the impact of noise correlations on coding depends on how they relate to neurons' stimulus preferences \cite{abbott_effect_1999, sompolinsky_population_2001, josic_stimulus-dependent_2009, ecker_effect_2011, moreno-bote_information-limiting_2014, hu_sign_2014,zylberberg2016direction,franke2016structures}.  In our study we explore a novel and complementary viewpoint on the impact of noise correlations on population coding.

\paragraph{}

Noise correlations are often related to neurons' stimulus tuning \cite{aertsen_dynamics_1989, ahissar_dependence_1992, espinosa_cortical_1988, kohn_stimulus_2005, ruff_attention_2014}---neuron pairs with similar tuning show larger noise correlations than pairs with dissimilar tuning.  The mechanisms behind noise correlations are varied \cite{doiron2016mechanics}, and both feedforward \cite{kanitscheider2015origin} and recurrent \cite{helias_correlation_2014} circuits can contribute to linking stimulus and noise correlations.  Further, in the absence of sensory stimulation, patterns of activity across cortical populations are often similar to those observed during sensory stimulation \cite{tsodyks_linking_1999, arieli_dynamics_1996, kenet_spontaneously_2003, luczak_spontaneous_2009, han_reverberation_2008, eagleman_image_2012, xu_activity_2012}.  Thus, the circuits that support correlated variability in spontaneous states likely overlap with the circuits responsible  for noise correlations in evoked states.  In other words, noise correlations may simply be a reflection of circuit dynamics that occur during periods when stimulus coding is not being performed.  

\paragraph{}

Spontaneous activity is usually viewed as a problem for plasticity: learned weight changes must be stable in the face of spontaneous activity.  Some previous studies have addressed this issue by endowing individual synapses with dynamical bistability between weak and strong weights \cite{graupner_stdp_2007, graupner_calcium-based_2012, higgins_memory_2014, zenke_diverse_2015}.  %These models potentially provide a rich dynamical explanation of the observation that individual CA3-CA1 synapses tend to occupy either strong or weak states and switch between them, rather than having a continuum of weights \cite{petersen_all-or-none_1998, oconnor_graded_2005}.  
%In contrast the STDP rule we have used here, as with other recently proposed models \cite{boustani_stable_2012, yger_convallis_2013}, relies on imposed upper and lower bounds for the synaptic weight to prevent synaptic weights from potentiating or depressing to unphysiological values.
By contrast, in our study the trained network architecture produced sizable within-assembly spontaneous correlations that combined with the STDP rule to reinforce assembly structure.  Thus, spontaneous activity did not dissolve learned architecture but rather preserved trained wiring.  If the assembly wiring was originally due to a shared stimulus input, then the spontaneous correlations needed to retain structure will be a source of noise correlations when the stimulus is to be coded.             

\paragraph{}  

Our simplified stimulus coding scenario was such that within-assembly noise correlations degraded the neural code.  However, the strong positive feedback from within-assembly recurrence enhanced the response gain, which improved coding.  Many studies of population coding separate response gain and response variability and for the purposes of analysis they are conceived as independent from one another \cite{abbott_effect_1999, sompolinsky_population_2001,ecker_effect_2011}.  While these studies have given insight into population coding, relating noise correlations and response gain to one another complicates analysis significantly \cite{series2004tuning,josic_stimulus-dependent_2009,zylberberg2016direction,franke2016structures}.  Our study expands on this general idea so that noise correlations are a reflection of the active maintenance of assembly structure and the high response gain it confers to a neuronal population.  This finding does not critically depend on the fast-timescale coordinated spiking activity required for STDP, and stability of assembly structure through long-timescale firing rate correlations should have a similar effect \cite{litwin-kumar_formation_2014,wei2014long}.

\section{Methods}
\subsection{Network model} \label{sec:network}
\paragraph{}
We consider a network of $N$ neurons, $N_E$ of which are excitatory and divided into $M$ clusters of size $\kappa$.  There are $N_I = \gamma \kappa$ inhibitory neurons.  Model parameters are in Table \ref{tab:par2}.  We take the excitatory-excitatory block of the adjacency matrix $\mathbf{W}^0$ to be partially symmetric: $\mathbf{W}^0_{EE} = \left(p_0-p_{0\text{sym}} \right) \mathbf{W}^0_\text{ER} + p_{0\text{sym}} \mathbf{W}^0_\text{sym}$ where $\mathbf{W}^0_\text{ER}$ has (directed, i.e. non-symmetric) Erd\H{o}s-R\'enyi statistics and $\mathbf{W}^0_\text{sym}$ is symmetric with Erd\H{o}s-R\'enyi statistics (i.e., as in an undirected graph).  Additionally we exclude autapses ($\mathbf{W}^0_{ii} = 0 $ $\forall i \in 1,\ldots,N$).

\paragraph{}
This means that the excitatory-excitatory connectivity is characterized by its empirical connection density $p_0$ and the frequency of loops $q_0$
\begin{equation} \begin{aligned}
p_0 &= \frac{1}{N_E^2}\sum_{i,j=1}^{N_E} \mathbf{W}^0_{ij} \\
q_0 &= \frac{1}{N_E^2}\sum_{i,j=1}^{N_E}\mathbf{W}^0_{ij}\mathbf{W}^0_{ji} - p_0^2
\end{aligned} \end{equation}

\begin{table}[!ht]
\caption{ \small
\bf{Model parameters}}
\begin{tabular}{|l|c|r|}
\hline
Parameter & Description & Value \\ \hline
$C$ & Membrane capacitance & 1 $\mu\mathrm{F/cm}^2$ \\ \hline
$g_L$ & Leak conductance & $0.1\mathrm{ mS/cm}^2$ \\ \hline
$V_L$ & Leak reversal potential & -72 mV \\ \hline
$\Delta$ & Action potential steepness & 1.4 mV\\ \hline
$V_T$ & Action potential initiation threshold & -48 mV \\ \hline
$V_{th}$ & Action potential threshold & 30 mV \\ \hline
$V_{re}$ & Action potential reset & -72 mV \\ \hline
$\tau_{ref}$ & Action potential width & 2 ms \\ \hline
$\mu$ & External input mean & 1 $\mu\mathrm{A/cm}^2$ \\ \hline
$\sigma$ & External input standard deviation & 9 mV \\ \hline
$W^\mathrm{max,E}$ & Maximum synaptic weight & $15\epsilon$ $\mu\mathrm{A/cm}^2$ \\ \hline
$W^\mathrm{max,I}$ & Maximum synaptic weight & $-7.5\epsilon$ $\mu\mathrm{A/cm}^2$ \\ \hline
$J(t)$ & Synaptic filter (EPSC shape) & $\exp{-(t / \tau_s)}$ \\ \hline
$\tau_\mathrm{sE}$ & Excitatory synaptic time constant & 2 ms \\ \hline
$\tau_\mathrm{sI}$ & Inhibitory time constant & 10 ms \\ \hline
\end{tabular}
\begin{flushleft}
\end{flushleft}
\label{tab:par2}
 \end{table}

\paragraph{}
We assume that the statistics of the adjacency matrix for within- and between-assembly connectivity are the same (and equal to $p_0$ and $q_0$).  The synaptic weight matrix, $\mathbf{W}$, is initially generated from $\mathbf{W}^0$ by giving each synapse the same initial weight.  We consider the mean strength of E-E synapses within one cluster $A$ and from other clusters into cluster $A$, $p_{AA}$ and $p_{AB}$ respectively:
\begin{equation} \begin{aligned}
\epsilon p_{AA} &= \frac{1}{\kappa^2}\sum_{i,j \in A} \mathbf{W}_{ij} \\
\epsilon p_{AB} &= \frac{1}{\kappa(N_E-\kappa)}\sum_{i \in A} \sum_{j \not\in A} \mathbf{W}_{ij}
\end{aligned} \end{equation}
The small parameter $\epsilon = (\kappa p_0)^{-1}$ scales the synaptic weights.  We take the mean strength of connections within each cluster to be symmetric and the strength of connections into any one cluster from outside to be the same as into the others (so for all clusters $A$ and $B$, $p_{AA} = p_{BB}$ and $p_{AB}=p_{BA}$).  Similarly, we measure the strength of \emph{reciprocal} connections within a cluster, $q_{AA}$, or between clusters, $q_{AB}$:
\begin{equation} \begin{aligned}
\epsilon q_{AA} &= \frac{1}{\kappa^2}\sum_{i,j \in A}\mathbf{W}_{ij}\mathbf{W}^0_{ji} - \epsilon p_0p_{AA} \\
\epsilon q_{AB} &= \frac{1}{\kappa(N_E-\kappa)}\sum_{i\in A}\sum_{j\not\in A}\mathbf{W}_{ij}\mathbf{W}^0_{ji} - \epsilon p_0p_{AB} \\
\end{aligned} \end{equation}
By subtracting off $p_0p_{AA}$ in the definition of $q_{AA}$ (and likewise for $q_{AB}$), we measure the mean strength of reciprocal connections above what would be expected in a network with no correlations between synapses.  Note: if the network is asymmetric ($\mathbf{W}^0_\text{sym} = \mathbf{0}$) then $q_0$ is negligible ($\mathcal{O}(\epsilon^{-3/2})$) and so are the initial values of $q_{AA}$ and $q_{AB}$.

\paragraph{}
We take the connectivity in between inhibitory and excitatory neurons, and within inhibitory neurons, to have (asymmetric) Erd\H{o}s-R\'enyi statistics, so that these are characterized by their mean synaptic weights: $p_{EI}$ for inhibitory $\rightarrow$ excitatory connections,
\begin{equation}
\epsilon p_{EI} = \frac{1}{N_E N_I}\sum_{i=1}^{N_E}\sum_{j=N_E+1}^N \mathbf{W}_{ij},
\end{equation}
and likewise $p_{IE}$ and $p_{II}$.

\paragraph{}
Finally, individual neurons had exponential integrate-and-fire (EIF) dynamics \cite{fourcaud-trocme_how_2003}, part of a class of models well-known to capture the spike initiation dynamics of cortical neurons \cite{jolivet_generalized_2004, jolivet_quantitative_2008}. Neurons' membrane voltages obeyed:
\begin{equation} \label{dVdt}
C\frac{dV_i}{dt} = g_L\left(V_L-V_i \right) + g_L\Delta\exp{\left(\frac{V_i-V_T}{\Delta}\right)} + I_\mathrm{i}(t) + \sum_{j=1}^N \mathbf{W}_{ij}\left(J(t)*y_j(t).\right).
\end{equation}
with model parameters in Table \ref{tab:par2}.

\subsection{Plasticity models} \label{sec:plast}
\paragraph{}
Synapses between excitatory neurons undergo additive Hebbian STDP:
\begin{equation}
\epsilon L(s)=\begin{cases} \mathcal{H}(W^\text{max}-\mathbf{W}_{ij})f_+e^{-\frac{\left| s \right|}{\tau_+}}, &\text{if } s \geq 0 \\
\mathcal{H}(\mathbf{W}_{ij}) \left(-f_-\right)e^{-\frac{\left| s\right|}{\tau_-}}, &\text{if } s < 0,
\end{cases}.
\end{equation}
where $s = t_{post}-t_{pre}$ is the time lag between spikes.  $f_\pm$ give the amplitude of individual changes in synaptic weights due to potentiation $(f_+)$ or depression $(f_-)$, and the time constants $\tau_\pm$ determine how synchronous spike pairs must be to cause plasticity.  When $f_\pm \ll W^\text{max}$, so that the timescale of plasticity is much longer than that of the STDP rule, individual weights undergo diffusion \cite{kempter_hebbian_1999} and their drift can be calculated as:
\begin{equation} \label{dwdt}
\frac{d\mathbf{W}_{ij}}{dt} = \mathbf{W}^0_{ij}\int_{-\infty}^\infty \epsilon L(s)\big(r_ir_j + \mathbf{C}_{ij}(s)\big)ds.
\end{equation}
Here, $r_i$ is the time-averaged firing rate of neuron $i$ and $\mathbf{C}_{ij}(s)$ is the spike train cross-covariance function of neurons $i$ and $j$.  We will assume that the integral of $L(s)$ is small enough ($\mathcal{O}(\epsilon)$) so that firing rates do not dominate the plasticity.

\paragraph{}
The inhibitory STDP rule is
\begin{equation} \label{Istdprule}
\epsilon L_I(s)=\mathcal{H}(\mathbf{W}_{ij}-W^\text{max,I}) f_I e^{-\frac{\left| s\right|}{\tau_I}}.
\end{equation}
\paragraph{}
In addition to this pair-based rule, each presynaptic (inhibitory) spike drives depression of the inhibitory synapses by $\mathcal{H}(-\mathbf{W}_{ij}) d_I = -2f_I \bar{r}_E\tau_I$.  This gives inhibitory $\rightarrow$ excitatory synapses a drift of
\begin{equation} \label{dwIdt}
\frac{d\mathbf{W}_{ij}}{dt} = \mathbf{W}^0_{ij}\Big(\int_{-\infty}^\infty \epsilon L_I(s)\big(r_ir_j + \mathbf{C}_{ij}(s)\big)ds - 2f_I\tau_I \bar{r}_Er_j\Big).
\end{equation}

\subsection{Derivation of assembly dynamics} \label{sec:deriv}
\paragraph{}
Here we will derive the dynamics of the assembly structure in networks of integrate-and-fire neurons undergoing STDP.  We will begin by considering the dynamics of mean synaptic weights and mean reciprocal synaptic weights both within and between assemblies.  The dynamics of $(p,q)$ and $(p_\Delta,q_\Delta)$ considered in the main text will then be recovered at the end.  The derivation follows the same steps as the derivation of the motif dynamics in \cite{ocker_self-organization_2015}.  We begin by expanding the covariance matrix $\mathbf{C}$ in path lengths through the network \cite{pernice_how_2011, trousdale_impact_2012} and truncating at first order in the interactions to obtain:
\begin{equation} \begin{aligned} \label{Ctrunc_inh}
\mathbf{C}_{ij}(s) &\approx  \overbrace{\delta_{ij} \mathbf{C}^0_{ij}(s)}^{\text{autocovariance}} + \overbrace{\left(\mathbf{A}_i*\mathbf{C^\eta}*\mathbf{A}_j\right)(s)}^{\text{external inputs}} + \overbrace{\left(\mathbf{W}_{ij}\mathbf{K}_{ij}*\mathbf{C}^0_{jj}\right)(s)}^{\text{forwards connections}} + \overbrace{\left(\mathbf{C}^0_{ii}*\mathbf{W}_{ji}\mathbf{K}_{ji}^-\right)(s)}^{\text{backwards connections}} \\
&\;\;\; + \underbrace{\sum_{k=1}^{N_E}\left(\mathbf{W}_{ik}\mathbf{K}_{ik}*\mathbf{C}^0_{kk}*\mathbf{W}_{jk}\mathbf{K}_{jk}^- \right)(s)}_{\text{common E inputs}} + \underbrace{\sum_{k=N_E+1}^{N}\left(\mathbf{W}_{ik}\mathbf{K}_{ik}*\mathbf{C}^0_{kk}*\mathbf{W}_{jk}\mathbf{K}_{jk}^- \right)(s)}_{\text{common I inputs}}.
\end{aligned} \end{equation}

\paragraph{}
As can be seen from Eqs. \eqref{dwdt} and \eqref{dwIdt}, these cross-covariances will control plasticity through their integral against the STDP rule.  We define variables measuring these STDP-weighted covariances (factoring out their amplitude, given by the $\mathbf{W}_{ij}$ factors in Eq. \eqref{Ctrunc_inh}):
\begin{equation} \begin{aligned}
S &= \int_{-\infty}^\infty L(s) ds \\ 
S_\eta &= \int_{-\infty}^\infty L(s)\left(A_E(t)*A_E(-t)\right)ds \\
S_F &= \int_{-\infty}^\infty L(s)\left(K_{EE}(t)*C_E^0(s) \right)ds \\
S_B &= \int_{-\infty}^\infty L(s)\left(C_E^0(s)*K_{EE}(-t) \right)ds \\
S_C &= \int_{-\infty}^\infty L(s)\left(K_{EE}(t)*C_E^0(s)*K_{EE}(-t) \right)ds \\
S_C^I &= \int_{-\infty}^\infty L(s)\left(K_{EI}(t)*C_I^0(s)*K_{EI}(-t) \right)ds 
\end{aligned} \end{equation}
and
\begin{equation} \begin{aligned}
S^I &= \int_{-\infty}^\infty L_I(s)ds = 2f_I\tau_I \\
S^{EI}_\eta &= \int_{-\infty}^\infty L_I(s)\big(A_E(t)*A_I(-t) \big) ds \\
S^{EI}_F &= \int_{-\infty}^\infty L_I(s)\big(K_{EI}(t)*C^0_I(s)\big) ds \\
S^{EI}_B &= \int_{-\infty}^\infty L_I(s)\big(C^0_E(s)*K_{IE}(-t)\big)ds \\
S^{EIE}_C &= \int_{-\infty}^{\infty} L_I(s)\big(K_{EE}(t)*C^0_E(s)*K_{IE}(-t)\big)ds \\
S^{EII}_C &= \int_{-\infty}^{\infty} L_I(s)\big(K_{EI}(t)*C^0_I(s)*K_{II}(-t) \big)ds.
\end{aligned} \end{equation}
In each of these definitions, $A_\alpha(t)$ corresponds to the mean linear response function of neurons of type $\alpha$, $\alpha \in \{E,I\}$.  $K_{\alpha\beta}(t)$ is the convolution of $A_\alpha(t)$ and the synaptic filter for synapses from $\beta$ neurons to $\alpha$ neurons ($\alpha, \beta \in \{E,I\}$).  We also define $r_E$ and $r_I$, the average excitatory and inhibitory firing rates.  Note that each of these are implicitly functions of the mean synaptic drive onto excitatory and inhibitory neurons.  Note that for the iSTDP rule, each presynaptic spike causes depression by $-S^I \bar{r}_E$.

\paragraph{}
We want the dynamics of the connectivity variables $p_{AA},p_{AB},q_{AA},q_{AB}$, so we differentiate these with respect to time.  Then, inserting Eq. \eqref{Ctrunc_inh} into Eq. \eqref{dwdt} and this into $dp_{AA}/dt$ yields:
\begin{equation} \begin{aligned} \label{dpAAdt_sums}
\frac{dp_{AA}}{dt} &= \left( r_E^2 S + c_{AA}\sigma^2 S_\eta\right) \frac{1}{\kappa^2}\sum_{i,j \in A} \mathbf{W}^0_{ij} + S_F  \frac{1}{\kappa^2}\sum_{i,j \in A} \mathbf{W}^0_{ij}\mathbf{W}_{ij} + S_B  \frac{1}{\kappa^2}\sum_{i,j\in A} \mathbf{W}^0_{ij}\mathbf{W}_{ji} \\ 
&\;\;\; + S_C  \frac{1}{\kappa^2}\sum_{i,j\in A}\sum_{k=1}^{N_E} \mathbf{W}^0_{ij}\mathbf{W}_{ik}\mathbf{W}_{jk} + S_C^I  \frac{1}{\kappa^2}\sum_{i,j\in A} \sum_{k=N_E+1}^N \mathbf{W}^0_{ij} \mathbf{W}_{ik}\mathbf{W}_{jk}
\end{aligned} \end{equation}
and similar for $p_{AB}$:
\begin{equation} \begin{aligned}
\frac{dp_{AB}}{dt} &= \left( r_E^2 S + c_{AB}\sigma^2 S_\eta\right) \frac{1}{\kappa(N_E-\kappa)}\sum_{i \in A} \sum_{j \not\in A} \mathbf{W}^0_{ij} + S_F  \frac{1}{\kappa(N_E-\kappa)}\sum_{i \in A} \sum_{j \not\in A} \mathbf{W}^0_{ij}\mathbf{W}_{ij} \\
&\;\;\; + S_B\frac{1}{\kappa(N_E-\kappa)}\sum_{i \in A} \sum_{j \not\in A} \mathbf{W}^0_{ij}\mathbf{W}_{ji} + S_C \frac{1}{\kappa(N_E-\kappa)}\sum_{i \in A} \sum_{j \not\in A} \sum_{k=1}^{N_E} \mathbf{W}^0_{ij}\mathbf{W}_{ik}\mathbf{W}_{jk} \\
&\;\;\; + S_C^I  \frac{1}{\kappa(N_E-\kappa)}\sum_{i \in A} \sum_{j \not\in A} \sum_{k=N_E+1}^N \mathbf{W}^0_{ij} \mathbf{W}_{ik}\mathbf{W}_{jk}.
\end{aligned} \end{equation}

\paragraph{}
The mean bidirectional connection strengths similarly evolve according to:
\begin{equation} \begin{aligned}
\frac{dq_{AA}}{dt} &= \left( r_E^2 S + c_{AA}\sigma^2 S_\eta\right) \frac{1}{\kappa^2}\sum_{i,j\in A} \mathbf{W}^0_{ij}\mathbf{W}^0_{ji} + S_F  \frac{1}{\kappa^2}\sum_{i,j\in A} \mathbf{W}^0_{ij}\mathbf{W}_{ij}\mathbf{W}^0_{ji} + S_B  \frac{1}{\kappa^2}\sum_{i,j\in A} \mathbf{W}^0_{ij}\mathbf{W}_{ji}\mathbf{W}^0_{ji} \\ 
&\;\;\; + S_C  \frac{1}{\kappa^2}\sum_{i,j\in A}\sum_{k=1}^{N_E} \mathbf{W}^0_{ij}\mathbf{W}_{ik}\mathbf{W}_{jk}\mathbf{W}^0_{ji} + S_C^I  \frac{1}{\kappa^2}\sum_{i,j\in A} \sum_{k=N_E+1}^N \mathbf{W}^0_{ij} \mathbf{W}_{ik}\mathbf{W}_{jk}\mathbf{W}^0_{ji} - p_0\frac{dp_{AA}}{dt}
\end{aligned} \end{equation}

\begin{equation} \begin{aligned}
\frac{dq_{AB}}{dt} &= \left( r_E^2 S + c_{AB}\sigma^2 S_\eta\right) \frac{1}{\kappa(N_E-\kappa)}\sum_{i \in A} \sum_{j \not\in A} \mathbf{W}^0_{ij}\mathbf{W}^0_{ji} + S_F \frac{1}{\kappa(N_E-\kappa)}\sum_{i \in A} \sum_{j \not\in A} \mathbf{W}^0_{ij}\mathbf{W}_{ij}\mathbf{W}^0_{ji} \\
&\;\;\; + S_B\frac{1}{\kappa(N_E-\kappa)}\sum_{i \in A} \sum_{j \not\in A} \mathbf{W}^0_{ij}\mathbf{W}_{ji}\mathbf{W}^0_{ji} + S_C  \frac{1}{\kappa(N_E-\kappa)}\sum_{i \in A} \sum_{j \not\in A} \sum_{k=1}^{N_E} \mathbf{W}^0_{ij}\mathbf{W}_{ik}\mathbf{W}_{jk}\mathbf{W}^0_{ji} \\
&\;\;\; + S_C^I  \frac{1}{\kappa(N_E-\kappa)}\sum_{i \in A} \sum_{j \not\in A} \sum_{k=N_E+1}^N \mathbf{W}^0_{ij} \mathbf{W}_{ik}\mathbf{W}_{jk}\mathbf{W}^0_{ji} - p_0 \frac{dp_{AB}}{dt}.
\end{aligned} \end{equation}

The mean inhibitory-to-excitatory synaptic weight obeys:
\begin{equation*} \begin{aligned} 
\frac{dp_{EI}}{dt} &= \frac{1}{N_EN_I}\sum_{i=1}^{N_E}\sum_{j=N_E+1}^N \frac{d\mathbf{W}_{ij}}{dt} \\
&= \frac{1}{N_EN_I}\sum_{i=1}^{N_E}\sum_{j=N_E+1}^N \mathbf{W}^0_{ij}\Big(\int_{-\infty}^\infty \epsilon L_I(s)\big(r_ir_j + \mathbf{C}_{ij}(s)\big)ds - 2f_I\tau_I \bar{r}_Er_I\Big).
 \end{aligned} \end{equation*}
\paragraph{}
Inserting the first-order truncation of spike train covariances yields:
\begin{equation} \begin{aligned} \label{dpEIdt_sums}
\frac{dp_{EI}}{dt} &= \left(r_I \big(r_E-\bar{r}_E\big)S^I +c_{EI}\sigma^2 S^{EI}_\eta   \right) \frac{1}{ N_EN_I} \sum_{i=1}^{N_E}\sum_{j=N_E+1}^N \mathbf{W}^0_{ij} \\
&\;\;\; +  S_F^{EI} \frac{1}{ N_EN_I}\sum_{i=1}^{N_E}\sum_{j=N_E+1}^N \mathbf{W}^0_{ij}\mathbf{W}_{ij} +  S^{EI}_B \frac{1}{ N_EN_I}\sum_{i=1}^{N_E}\sum_{j=N_E+1}^N \mathbf{W}^0_{ij}\mathbf{W}_{ji} \\
&\;\;\; + S^{EIE}_C \frac{1}{ N_EN_I}\sum_{i=1}^{N_E}\sum_{j=N_E+1}^N\sum_{k=1}^{N_E} \mathbf{W}^0_{ij} \mathbf{W}_{ik}\mathbf{W}_{jk} +  S^{EII}_C\frac{1}{ N_EN_I}\sum_{i=1}^{N_E}\sum_{j=N_E+1}^N\sum_{k=N_E+1}^{N} \mathbf{W}^0_{ij} \mathbf{W}_{ik}\mathbf{W}_{jk}.
\end{aligned}\end{equation}

\paragraph{}
The next step in writing down dynamics for each of the $p$ and $q$ variables of interest is to evaluate the sums over $\mathbf{W}$ and $\mathbf{W}^0$ in Eqs. \eqref{dpAAdt_sums}--\eqref{dpEIdt_sums}.  Recalling that the adjacency matrix is Erd\H{o}s-R\'enyi except for the partial symmetry of the excitatory-excitatory block, this yields (neglecting higher-order motif contributions):
\begin{equation}
\frac{dp_{AA}}{dt} = \left( r_E^2 S + c_{AA}\sigma^2 S_\eta\right) p_0 + S_F  \epsilon p_{AA} + S_B \epsilon \left(q_{AA}+p_0p_{AA}\right)+ S_C \epsilon^2 p_0  \left(\kappa p_{AA}^2 + \left(N_E-\kappa\right)p_{AB}^2 \right) + S_C^I \epsilon^2  N_I p_0 p_{EI}^2
\end{equation}
\begin{equation} \begin{aligned}
\frac{dp_{AB}}{dt} &= \left( r_E^2 S + c_{AB}\sigma^2 S_\eta\right) p_0 + S_F \epsilon p_{AB} + S_B \epsilon (q_{AB}+p_0p_{AB}) \\
&\;\;\; + S_C \epsilon^2 p_0 \left(2\kappa p_{AA}p_{AB} + \left(N_E-2\kappa \right)p_{AB}^2 \right) 
+ S_C^I \epsilon^2 N_I p_0 p_{EI}^2
\end{aligned} \end{equation}
\begin{equation} \begin{aligned}
\frac{dq_{AA}}{dt} &= \left( r_E^2 S + c_{AA}\sigma^2 S_\eta\right)q_0 +  S_F \epsilon q_{AA} + S_B \epsilon(1-p_0)(q_{AA}+p_0p_{AA}) \\ 
&\;\;\; + S_C \epsilon^2 q_0 \left(\kappa p_{AA}^2+(N_E-\kappa)p_{AB}^2\right)+ S_C^I \epsilon^2 N_I q_0p_{EI}^2
\end{aligned} \end{equation}
\begin{equation} \begin{aligned}
\frac{dq_{AB}}{dt} &= \left( r_E^2 S + c_{AB}\sigma^2 S_\eta\right)q_0 + S_F\epsilon q_{AB} + S_B \epsilon \left(1-p_0\right)\left(q_{AB}+p_0p_{AB}\right) \\
&\;\;\; + S_C \epsilon^2 q_0\left(2\kappa p_{AA}p_{AB}+\left(N_E-2\kappa \right)p_{AB}^2\right) + S_C^I \epsilon^2q_0N_Ip_{EI}^2
\end{aligned} \end{equation}
\begin{equation} \begin{aligned}
\frac{dp_{EI}}{dt} &= \left(r_I \big(r_E-\bar{r}_E\big)S^I +c_{EI}\sigma^2 S^{EI}_\eta \right)p_0^{EI} + S_F^{EI}\epsilon p_{EI} + S_B^{EI} \epsilon p_0^{EI}p_{IE} \\
&\;\;\; + S_C^{EIE} \epsilon^2 p_0^{EI} p_{IE}\left(\kappa p_{AA} + \left(N_E-\kappa\right)p_{AB} \right) + S_C^{EII} \epsilon^2 p_0^{EI}N_I p_{EI}p_{II}.
\end{aligned} \end{equation}

\paragraph{}
Finally, we recall that $\epsilon = (\kappa p_0)^{-1}$ and $N_I = \gamma \kappa$, revealing that the dynamics above stop at $\mathcal{O}(\epsilon)$:
\begin{equation} 
\frac{dp_{AA}}{dt} = \left( r_E^2 S + c_{AA}\sigma^2 S_\eta\right) p_0 + \epsilon \left[S_F  p_{AA} + S_B \left(q_{AA}+p_0p_{AA}\right)+ S_C  \left(p_{AA}^2 + \left(M-1\right)p_{AB}^2 \right) + S_C^I  \gamma p_{EI}^2 \right]
\end{equation}

\begin{equation} \begin{aligned}
\frac{dp_{AB}}{dt} = \left( r_E^2 S + c_{AB}\sigma^2 S_\eta\right) p_0 + \epsilon & \big[ S_F p_{AB} 
+ S_B(q_{AB}+p_0p_{AB})+ S_C \left(2p_{AA}p_{AB} + \left(M-2 \right)p_{AB}^2 \right) \\
& + S_C^I  \gamma p_{EI}^2 \big]
\end{aligned} \end{equation}

\begin{equation} \begin{aligned}
\frac{dq_{AA}}{dt} = \left( r_E^2 S + c_{AA}\sigma^2 S_\eta\right)q_0 + \epsilon & \big[S_F q_{AA} + S_B(1-p_0)(q_{AA}+p_0p_{AA}) \\ 
&\;\;\; + S_C \frac{1}{p_0} q_0 \left(p_{AA}^2+(M-1)p_{AB}^2\right)+ S_C^I  \frac{\gamma}{p_0}q_0p_{EI}^2 \big]
\end{aligned} \end{equation}

\begin{equation} \begin{aligned}
\frac{dq_{AA}}{dt} = \left( r_E^2 S + c_{AA}\sigma^2 S_\eta\right)q_0 + \epsilon & \big[S_F q_{AA} + S_B(1-p_0)(q_{AA}+p_0p_{AA}) \\ 
&\;\;\; + S_C \frac{1}{p_0} q_0 \left(p_{AA}^2+(M-1)p_{AB}^2\right)+ S_C^I  \frac{\gamma}{p_0}q_0p_{EI}^2 \big]
\end{aligned} \end{equation}

\begin{equation} \begin{aligned}
\frac{dq_{AB}}{dt} = \left( r_E^2 S + c_{AB}\sigma^2 S_\eta\right)q_0 + \epsilon &\big[ S_Fq_{AB} + S_B\left(1-p_0\right)\left(q_{AB}+p_0p_{AB}\right) \\
&\;\;\; + S_C \frac{1}{p_0}q_0\left(2p_{AA}p_{AB}+\left(M-2\right)p_{AB}^2\right) + S_C^I  \frac{\gamma}{p_0}q_0p_{EI}^2 \big]
\end{aligned} \end{equation}

\begin{equation} \begin{aligned}
\frac{dp_{EI}}{dt} = \left(r_I \big(r_E-\bar{r}_E\big)S^I +c_{EI}\sigma^2 S^{EI}_\eta \right)p_0^{EI} + \epsilon & \big[S_F^{EI} p_{EI} + S_B^{EI}p_0^{EI}p_{IE} \\
&+ S_C^{EIE} \frac{p_0^{EI}}{p_0}p_{IE}\left(p_{AA} + \left(M-1\right)p_{AB} \right) + S_C^{EII}\frac{p_0^{EI}}{p_0}\gamma p_{EI}p_{II} \big].
\end{aligned} \end{equation}

\subsection{Firing rate dynamics}
\paragraph{}
Here we have written the dynamics in terms of the average firing rates $r_E,r_I$ and STDP-weighted spiking covariances as if those were parameters.  As the mean excitatory and inhibitory weights change, so will neurons' firing rates.  We now supplement the dynamics of the connectivity by examining the evolution of the population-averaged firing rates $r_\alpha$ with $\alpha \in \{E,I\}$.  The quasi-stationary firing rates obey:
\begin{equation}
r_\alpha(t) = f_\alpha(\mu_\alpha(t),\sigma^2)
\end{equation}
where $f_\alpha$ is the rate-current function of an EIF neuron belonging to population $\alpha$ and
\begin{equation} \begin{aligned}
\mu_E & = \mu_{\text{ext},E} + \epsilon N_Ep\tau_Er_E + \epsilon N_I p_{EI}\tau_Ir_I \\
&= \mu_{\text{ext},E} + \epsilon \left(\kappa p_{AA} + (N_E-\kappa)p_{AB} \right)\tau_Er_E + \epsilon N_I p_{EI}\tau_Ir_I \\
&= \mu_{\text{ext},E} + \frac{1}{p_0}\left(p_{AA} + \left(M-1\right)p_{AB}\right)\tau_Er_E + \frac{\gamma}{p_0} p_{EI}\tau_Ir_I  \\
\mu_I &= \mu_{\text{ext},I} + \epsilon N_Ep_{IE}\tau_Er_E + \epsilon N_I p_{II}\tau_Ir_I \\
&= \mu_{\text{ext},I} + \frac{M}{p_0} p_{IE}\tau_Er_E + \frac{\gamma}{p_0} p_{II}\tau_Ir_I
\end{aligned} \end{equation}
 is the average external input to one of those neurons and we assume that a sufficient combination of low firing rates and weak/slow synapses keeps recurrent connectivity from contributing significantly to the effective variance of inputs to a neuron. $\tau_E$ and $\tau_I$ are the integrals of excitatory and inhibitory synaptic kernels (these are described by single exponentials, so the integral is their decay time constant).

\paragraph{}
The dynamics of the quasi-stationary firing rates is then given by:
\begin{equation}
\frac{dr_\alpha}{dt} = \frac{df_\alpha}{d\mu_\alpha}\frac{d\mu_\alpha}{dt}.
\end{equation}  
Recalling that $\left. \frac{df_\alpha}{d\mu}\right|_{\mu_\alpha} = \int_0^\infty A_\alpha(t) dt$, where $A_\alpha(t)$ is the average linear response of neurons of type  $\alpha$, we define
\begin{equation}
S^\alpha_A \equiv \int_0^\infty A_\alpha(t) dt.
\end{equation}
Assuming that $\mu_{\text{ext},\alpha}$ is constant in time, we obtain:
\begin{equation} \begin{aligned} \label{drEdt_drI}
\frac{dr_E}{dt} &= S^E_A \left(\frac{\tau_E}{p_0}\left(\left(p_{AA}+\left(M-1\right)p_{AB}\right)\frac{dr_E}{dt} + \left(\frac{dp_{AA}}{dt} + (M-1)\frac{dp_{AB}}{dt} \right)r_E \right) + \frac{\gamma \tau_I}{p_0}\left(\frac{dp_{EI}}{dt}r_I + p_{EI}\frac{dr_I}{dt} \right)\right)
\end{aligned} \end{equation}
and since the excitatory $\rightarrow$ inhibitory and inhibitory $\rightarrow$ inhibitory weights are not plastic, $r_I$ tracks $r_E$:
\begin{equation} \begin{aligned} \label{drIdt_drE}
\frac{dr_I}{dt} &= S^I_A \left(\frac{M}{p_0}p_{IE}\tau_E\frac{dr_E}{dt} + \frac{\gamma}{p_0}p_{II}\tau_I\frac{dr_I}{dt}\right) \\
&= \left(\frac{S^I_A \frac{M}{p_0}p_{IE}\tau_E}{1-\frac{\gamma}{p_0}p_{II}\tau_I}\right)\frac{dr_E}{dt}.
\end{aligned} \end{equation}
Inserting Eq. \eqref{drIdt_drE} into Eq. \eqref{drEdt_drI} then yields
\begin{equation}
\frac{dr_E}{dt} = \frac{S^E_A \left(\frac{\tau_E}{p_0}\left(\frac{dp_{AA}}{dt} + (M-1)\frac{dp_{AB}}{dt} \right)r_E + \frac{\gamma \tau_I}{p_0}\frac{dp_{EI}}{dt}r_I \right)}{\left(1- S^E_A\frac{\tau_E}{p_0}\left(p_{AA}+\left(M-1\right)p_{AB}\right) - S^E_A\frac{\gamma\tau_I}{p_0}p_{EI}\left(\frac{S^I_A \frac{M}{p_0}p_{IE}\tau_E}{1-\frac{\gamma}{p_0}p_{II}\tau_I}\right)\right)}.
\end{equation}

\subsection{Linear stability of firing rates} \label{sec:pEIunbalstab}
\paragraph{}
The iSTDP rule imposes a form of rate homeostasis on the dynamics, keeping $r_E$ within $\mathcal{O}(\epsilon)$ of $\bar{r}_E$.  Indeed, this was one major motivation for its theoretical proposal (Sprekeler \& Vogels et al, 2011).  We now check how this affects the dynamics of the weights.  If there is a balance between potentiation and depression in the eSTDP rule $L(s)$ so that $S \sim \mathcal{O}(\epsilon)$, then the dynamics of mean excitatory weights have an $\mathcal{O}(1/\epsilon)$ timescale.  There is a different condition for balance between potentiation and depression of inhibitory $\rightarrow$ excitatory synapses.  This balance occurs when the excitatory rate is close to $\bar{r}_E$, requiring $(r_E-\bar{r}_E) \sim \mathcal{O}(\epsilon)$.  If the eSTDP rule is balanced but $(r_E-\bar{r}_E) \sim \mathcal{O}(1)$ then the leading order dynamics of the firing rates and $p_{EI}$ become $\mathcal{O}(1)$ and obey Eq. \eqref{dpEIdt_unbal}:
%\begin{equation} \label{dpEIdt_unbal}
%\frac{dp_{EI}}{dt} = \left(r_I \big(r_E-\bar{r}_E\big)S^I +c_{EI}\sigma^2 S^{EI}_\eta \right)p_0^{EI} 
%\end{equation}
\begin{equation} \label{drEdt_unbal}
\frac{dr_E}{dt} = \underbrace{\left(\frac{S^E_A \frac{\gamma \tau_I}{p_0}}{1- S^E_A\frac{\tau_E}{p_0}\left(p_{AA}+\left(M-1\right)p_{AB}\right) - S^E_A\frac{\gamma\tau_I}{p_0}p_{EI}\left(\frac{S^I_A \frac{M}{p_0}p_{IE}\tau_E}{1-\frac{\gamma}{p_0}p_{II}\tau_I}\right)}\right)}_{X(p_{EI})}r_I\frac{dp_{EI}}{dt}
\end{equation}
\begin{equation} \label{drIdt_unbal}
\frac{dr_I}{dt} = \underbrace{\left(\frac{S^I_A \frac{M}{p_0}p_{IE}\tau_E}{1-\frac{\gamma}{p_0}p_{II}\tau_I}\right)}_{Y}\frac{dr_E}{dt}
\end{equation}
with fixed points $(p_{EI}^*,r_I^*,r_E^*)$ obeying:
\begin{equation} \begin{aligned}
0 &= \left(r_I^*\big(r_E^*-\bar{r}_E\big)S^I + c_{EI}\sigma^2 S^{EI}_\eta \right) \\
0 &= X(p_{EI}^*)\cdot r_I^* \cdot \left(r_I^* \big(r_E^*-\bar{r}_E\big)S^I + c_{EI}\sigma^2 S^{EI}_\eta \right) \\
0 &= Y \cdot X(p_{EI}^*)\cdot r_I^* \cdot \left(r_I^* \big(r_E^*-\bar{r}_E\big)S^I + c_{EI}\sigma^2 S^{EI}_\eta \right).
\end{aligned} \end{equation}
In order for the first condition to hold ($dp_{EI}/dt=0$), the fixed point rates must lie on the hyperbola given by 
\begin{equation}
r_E^* = -\frac{c_{EI}\sigma^2 S^{EI}_\eta}{S^I}\left(\frac{1}{r_I^*}\right) + \bar{r}_E .
\end{equation}
This also satisfies $dr_{E}/dt= 0$ and $dr_I/dt = 0$.  If $c_{EI} = 0$, this reduces to $r_E^* = \bar{r}_E$.

\paragraph{}
We next examine the linear stability of this solution.  The Jacobian for Eqs. \eqref{dpEIdt_unbal}, \eqref{drEdt_unbal}--\eqref{drIdt_unbal} is:
\[\left(
\begin{array}{ccc}
0 & r_IS^Ip_0^{EI} & (r_E-\bar{r}_E)S^Ip_0^{EI} \\
r_I\left(r_I \big(r_E-\bar{r}_E\big)S^I +c_{EI}\sigma^2 S^{EI}_\eta \right)p_0^{EI} \frac{\partial X}{\partial p_{EI}} & X r_I^2S^Ip_0^{EI} & 2Xp_0^{EI}(r_E-\bar{r}_E)S^Ir_I + Xc_{EI}\sigma^2S_\eta^{EI} \\
Y r_I\left(r_I \big(r_E-\bar{r}_E\big)S^I +c_{EI}\sigma^2 S^{EI}_\eta \right)p_0^{EI} \frac{\partial X}{\partial p_{EI}} & Y X r_I^2S^Ip_0^{EI} & 2YXp_0^{EI}(r_E-\bar{r}_E)S^Ir_I + YXc_{EI}\sigma^2S_\eta^{EI}
\end{array} 
\right) \]
where
\begin{equation}
\frac{\partial X}{\partial p_{EI}} = \frac{\left(S^E_A \frac{\gamma \tau_I}{p_0}\right)^2\left(\frac{S^I_A \frac{M}{p_0}p_{IE}\tau_E}{1-\frac{\gamma}{p_0}p_{II}\tau_I}\right)}{\left(1- S^E_A\frac{\tau_E}{p_0}\left(p_{AA}+\left(M-1\right)p_{AB}\right) - S^E_A\frac{\gamma\tau_I}{p_0}p_{EI}\left(\frac{S^I_A \frac{M}{p_0}p_{IE}\tau_E}{1-\frac{\gamma}{p_0}p_{II}\tau_I}\right)\right)^2}
\end{equation}
The eigenvalues of the Jacobian, evaluated at $p_{EI}^*,r_E^* = \bar{r}_E,r_I^*$ with $c_{EI}=0$, are:
\begin{equation} \begin{aligned}
\lambda_1 &= \lambda_2 = 0, \\
\lambda_3 &= \frac{\gamma \left(r_I^*\right)^2S^E_A S^I \tau_I}{1-\frac{S^E_A\tau_E\left(p_{AA}+(M-1)p_{AB} \right)}{p_0}-\frac{\gamma M p_{EI} p_{IE} S^E_A S^I_A \tau_E \tau_I}{p_0^2-\gamma p_0p_{II}\tau_I}}
\end{aligned} \end{equation}
Below, we plot these eigenvalues (with $c_{EI}=0$) as a function of the total excitation $p_{AA}+(M-1)p_{AB}$ with $c_{EI}=0$ so that $r_E^* = \bar{r}_E$.  For each $p_{AA}+(M-1)p_{AB}$, we use bisection to find $p_{EI}^* \in [0,\text{W}_\text{max}^I]$ that minimizes $\left| (r_E-\bar{r}_E) \right|$ (for the particular cellular and network parameters used).  Fortunately, the inhibition is strong enough to achieve $r_E = \bar{r}_E$ - it would be possible for this not to be the case, for example with weak W$_\text{max}^I$.

\subsection{Final dynamics of network structure: mind your p's and q's} \label{sec:dynamicsrbar}
The above analysis of unbalanced iSTDP reveals that there is a $\mathcal{O}(\epsilon)$ neighborhood around $p_{EI}^*,\bar{r}_E,r_I^*$ which is attracting along those dimensions, so that $r_E = \bar{r}_E+\mathcal{O}(\epsilon), r_I = r_I^*+\mathcal{O}(\epsilon), p_{EI} = p_{EI}^*+\mathcal{O}(\epsilon)$.  (If $c_{EI}\neq 0$ then $\lambda_2\neq 0$ and the dynamics could be different, a potential subject for future study.)   Inserting these yields the following equations, up to $\mathcal{O}(\epsilon)$ and for balanced eSTDP (so $S \sim \mathcal{O}(\epsilon)$): 
\begin{equation} 
\frac{dp_{AA}}{dt} = \left( \bar{r}_E^2 S + c_{AA}\sigma^2 S_\eta\right) p_0 + \epsilon \left[S_F  p_{AA} + S_B \left(q_{AA}+p_0p_{AA}\right)+ S_C  \left(p_{AA}^2 + \left(M-1\right)p_{AB}^2 \right) + S_C^I  \gamma (p_{EI}^*)^2 \right]
\end{equation}
\begin{equation} \begin{aligned}
\frac{dp_{AB}}{dt} = \left( \bar{r}_E^2 S + c_{AB}\sigma^2 S_\eta\right) p_0 + \epsilon & \big[ S_F p_{AB} 
+ S_B(q_{AB}+p_0p_{AB})+ S_C \left(2p_{AA}p_{AB} + \left(M-2 \right)p_{AB}^2 \right) \\
& + S_C^I  \gamma (p_{EI}^*)^2 \big]
\end{aligned} \end{equation}
\begin{equation} \begin{aligned}
\frac{dq_{AA}}{dt} = \left( \bar{r}_E^2 S + c_{AA}\sigma^2 S_\eta\right)q_0 + \epsilon & \big[S_F q_{AA} + S_B(1-p_0)(q_{AA}+p_0p_{AA}) \\ 
&\;\;\; + S_C \frac{1}{p_0} q_0 \left(p_{AA}^2+(M-1)p_{AB}^2\right)+ S_C^I  \frac{\gamma}{p_0}q_0(p_{EI}^*)^2 \big]
\end{aligned} \end{equation}
\begin{equation} \begin{aligned}
\frac{dq_{AB}}{dt} = \left( \bar{r}_E^2 S + c_{AB}\sigma^2 S_\eta\right)q_0 + \epsilon &\big[ S_Fq_{AB} + S_B\left(1-p_0\right)\left(q_{AB}+p_0p_{AB}\right) \\
&\;\;\; + S_C \frac{1}{p_0}q_0\left(2p_{AA}p_{AB}+\left(M-2\right)p_{AB}^2\right) + S_C^I  \frac{\gamma}{p_0}q_0(p_{EI}^*)^2 \big].
\end{aligned} \end{equation}
Note that the location of $(\bar{r}_E,r_I^*,p_{EI}^*)$ depends on the net excitation, $p_{AA}+(M-1)p_{AB}$, and so will evolve on the slow timescale of the balanced eSTDP.  We compute the nullclines of these equations in asymmetric networks by bisection.  For example, for each $p_{AA}$ we find the $p_{AB}$ for which the homeostatic $p_{EI}^*$ associated with ($p_{AA},p_{AB}$) gives $dp_{AA}/dt=0$.

\subsection{Temporally symmetric eSTDP} \label{sec:tempsym}
\paragraph{}
When the timescales of potentiation and depression in the excitatory STDP rule are similar, $\tau_+ \sim \tau_- + \mathcal{O}(\epsilon)$, then the dynamics of the network structure simplify considerably.  Since the correlations from common inputs (both from excitatory and inhibitory neurons) are temporally symmetric around 0 lag, this makes $S_C, S_C^I, S_C^{EIE}, S_C^{EII} \sim \mathcal{O}(\epsilon)$.  The dynamics then reduce to:

\begin{equation} 
\frac{dp_{\alpha}}{dt} = \left( \bar{r}_E^2 S + c_{\alpha}\sigma^2 S_\eta\right) p_0 + \epsilon \left[S_F  p_{\alpha} + S_B \left(q_{\alpha}+p_0p_{\alpha}\right)\right]
\end{equation}

\begin{equation} \begin{aligned}
\frac{dq_{\alpha}}{dt} = \left( \bar{r}_E^2 S + c_{\alpha}\sigma^2 S_\eta\right)q_0 + \epsilon & \big[S_F q_{\alpha} + S_B(1-p_0)(q_{\alpha}+p_0p_{\alpha})\big]
\end{aligned} \end{equation}

for $\alpha = AA$ or $AB$.

\subsection{Separable dynamics of assembly formation and segregation} \label{sec:Deltapq}
\paragraph{}
The dynamics of the network structure simplify if we take a linear transformation of our $p$ and $q$ variables:
\begin{equation} \begin{aligned}
p &= \frac{Mp_{AA} + M(M-1)p_{AB}}{M^2} \\
q &= \frac{Mq_{AA}+M(M-1)q_{AB}}{M^2} \\
p_\Delta &= p_{AA} - p_{AB} \\
q_\Delta &= q_{AA} - q_{AB}.
\end{aligned} \end{equation}
The first two, $p,q$, measure the total mean synaptic weight and the mean weight of reciprocal connections overall in the network.  The second two measure the formation of structure.  The dynamics of these transformed variables are:

\begin{equation}
\frac{dp}{dt} = \left(\bar{r}_E^2S+c_{EE}\sigma^2S_\eta\right)p_0 + \epsilon \left[S_Fp + S_B\left(q + p_0p\right) + S_C M p^2 + S_C^I\gamma (p_{EI}^*)^2\right]
\end{equation}
\begin{equation}
\frac{dq}{dt} = \left(\bar{r}_E^2S+c_{EE}\sigma^2S_\eta\right)q_0 + \epsilon \left[S_Fq + S_B\left(1-p_0\right)\left(q + p_0p\right) + S_C M \frac{q_0}{p_0}p^2 + S_C^I \gamma \frac{q_0}{p_0} (p_{EI}^*)^2\right]
\end{equation}
\begin{equation}
\frac{dp_\Delta}{dt} = c_\Delta \sigma^2S_\eta p_0 + \epsilon \left[S_Fp_\Delta + S_B(q_\Delta + p_0p_\Delta) + S_Cp_\Delta^2 \right]
\end{equation}
\begin{equation}
\frac{dq_\Delta}{dt} = c_\Delta \sigma^2S_\eta q_0 + \epsilon \left[S_Fq_\Delta + S_B\left(1-p_0\right)\left(q_\Delta + p_0p_\Delta \right) + S_C\frac{q_0}{p_0}p_\Delta^2  \right]
\end{equation}
where $c_{EE}$ is defined, analogously to $p$, as the average correlation of external inputs and $c_\Delta = c_{AA} - c_{AB}$.  Here we see that the spontaneous dynamics of overall potentation/depression ($p,q$) are separable from the dynamics of structure formation ($p_\Delta, q_\Delta$).  

\paragraph{}
The nullclines are given by solving each equation for the steady state, and are:
\begin{equation} \begin{aligned}
p^* &= \frac{-\epsilon(S_F+p_0S_B) \pm \sqrt{\left(\epsilon (S_F+p_0S_B)\right)^2 - 4\epsilon S_C M \left(\left(\bar{r}_E^2 S + c_{EE}\sigma^2 S_\eta\right)p_0 + \epsilon(S_Bq^* + S_C^I \gamma (p_{EI}^*)^2) \right)}}{2\epsilon S_C M} \\
q^* &= -\frac{\left(\bar{r}_E^2 S + c_{EE}\sigma^2S_\eta\right)q_0 + \epsilon \left(S_B(1-p_0)p_0p^* +S_C M \frac{q_0}{p_0} (p^*)^2 + S_C^I \gamma \frac{q_0}{p_0} (p_{EI}^*)^2\right)}{\epsilon (S_F+(1-p_0)S_B)}
\end{aligned} \end{equation}
\begin{equation} \begin{aligned}
p_\Delta^* &= \frac{-\epsilon(S_F+p_0S_B) \pm \sqrt{\epsilon^2\left(S_F+p_0S_B\right)^2 - 4\epsilon S_C \left(c_\Delta\sigma^2S_\eta p_0 + \epsilon S_Bq_\Delta^*\right)}}{2\epsilon S_C} \\
q_\Delta^* &= -\frac{c_\Delta\sigma^2S_\eta q_0 + \epsilon \left(S_B(1-p_0)p_0p_\Delta^* + S_C\frac{q_0}{p_0}(p_\Delta^*)^2 \right)}{\epsilon\left(S_F+S_B(1-p_0)\right)}
\end{aligned} \end{equation}
In the spontaneous case ($c_{EE} = c_\Delta = 0$) and defining $S = -\delta \epsilon$ these simplify to:
\begin{equation} \begin{aligned}
p^* &= \frac{-(S_F+p_0S_B) \pm \sqrt{\left( (S_F+p_0S_B)\right)^2 - 4S_C M \left(-\bar{r}_E^2 \delta p_0 + S_Bq^* + S_C^I \gamma (p_{EI}^*)^2 \right)}}{2 S_C M} \\
q^* &= -\frac{-\bar{r}_E^2 \delta q_0 + S_B(1-p_0)p_0p^* +S_C M \frac{q_0}{p_0} (p^*)^2 + S_C^I \gamma \frac{q_0}{p_0} (p_{EI}^*)^2}{S_F+(1-p_0)S_B}
\end{aligned} \end{equation}
\begin{equation} \begin{aligned}
p_\Delta^* &= \frac{-(S_F+p_0S_B) \pm \sqrt{\left(S_F+p_0S_B\right)^2 - 4S_C \left(S_Bq_\Delta^*\right)}}{2 S_C} \\
q_\Delta^* &= -\frac{S_B(1-p_0)p_0p_\Delta + S_C\frac{q_0}{p_0}(p_\Delta^*)^2 }{S_F+(1-p_0)S_B} .
\end{aligned} \end{equation}

\subsection{Fisher Information} \label{methods: fisher}
\paragraph{}
We consider the linear Fisher information of an assembly's activity about a stimulus $\theta$:
\begin{equation}
FI_{A} \approx \left( \frac{dn_A}{d\theta}\right)^2 C_{AA}^{-1} = \left( \kappa T \frac{dr_A}{d\theta}\right)^2 C_{AA}^{-1}
\end{equation}
where $C_{AA}$ is the variance of an assembly's summed spike count in a window of $T=100$ ms. We compute it, using the length one approximation of spike train covariances \eqref{Ctrunc_inh} as \cite{cox_point_1980}:
\begin{equation} \begin{aligned}
C_{AA} &= \sum_{i,j \in A} \left( \int_{-T}^T \left(T-\left|s\right|) \mathbf{C}_{ij}(s)\right)ds \right) \\ 
&= \int_{-T}^T \left(T-\left|s\right| \right)\Big[\kappa C^0_{E}(s) + \kappa\left(\kappa-1\right)\Big(c_{AA}\sigma^2 \left(A_E(t)*A_E(-t)\right) + p_{AA}\left(K_{EE}*C^0_E(s)\right) \\
&\;\;\; + \left(q_{AA}+p_0p_{AA}\right)C^0_E(s)*K_{EE}(-t) + \left(\kappa p_{AA}^2 + (N_E-\kappa)p_{AB}^2 \right) \left(K_{EE}(t)*C^0_E(s)*K_{EE}(-t)\right) \\
&\;\;\; + N_I \left(p_{EI}^*\right)^2 \left(K_{EI}(t)*C^0_I(s)*K_{EI}(-t) \right)\Big)  \Big] ds
\end{aligned} \end{equation}

\paragraph{}
In order to calculate the stimulus-response gain $\frac{dr_A}{d\theta}$, we consider all sources of input to neurons in assembly $A$:
\begin{equation} \begin{aligned}
 \mu_\mathrm{tot,A} 
 %&= \mu_\mathrm{ext} + \mu_\theta +  \epsilon \kappa \tau_E p_{AA}r_{A} + \epsilon(N_E-\kappa) \tau_E p_{AB}r_B   + N_I \tau_I \epsilon p_{EI}r_I \\
 &= \mu_\mathrm{ext} + \mu_\theta + \frac{\tau_E}{p_0} p_{AA}r_A + \frac{\tau_E}{p_0}(M-1)p_{AB}r_B + \frac{\gamma \tau_I}{p_0}p_{EI}^* r_I
\end{aligned} \end{equation}
(using $\epsilon = (\kappa p_0)^{-1}$), where $r_A$ and $r_B$ are the rates of excitatory neurons in the stimulated $(r_A)$ or non-stimulated $(r_B)$ assemblies. The total inputs to excitatory neurons in non-stimulated assemblies and inhibitory neurons are:
\begin{equation} \begin{aligned}
\mu_\mathrm{tot,B} 
%&=  \mu_\mathrm{ext} + \epsilon \left( \kappa \tau_E p_{AA}r_{B} + \kappa \tau_E p_{AB}r_A +(N_E-2\kappa) \tau_E p_{AB}r_B   + N_I \tau_I p_{EI}r_I \right) \\
&=\mu_\mathrm{ext} + \frac{\tau_E}{p_0}\left(p_{AA}r_B + p_{AB}r_A + (M-2)p_{AB}r_B \right) + \frac{\gamma \tau_I}{p_0} p_{EI}^* r_I \\
\mu_\mathrm{tot,I} 
%&= \mu_\mathrm{ext} + \epsilon \tau_E p_{IE}N_E r_E + \epsilon N_I \tau_I p_{II} r_I \\
%&= \mu_\mathrm{ext} + \epsilon \tau_E p_{IE} \left(\kappa r_A + (N_E-\kappa)r_B \right)+ \epsilon N_I \tau_I p_{II} r_I \\
&= \mu_\mathrm{ext} + \frac{\tau_E}{p_0} p_{IE}\left(r_A + (M-1)r_B \right) + \frac{\gamma \tau_I}{p_0} p_{II}r_I.
\end{aligned} \end{equation}

\paragraph{}
Applying the chain rule gives:
\begin{equation} \begin{aligned} \label{drA_non_isolated}
\frac{dr_A}{d\theta} &= \frac{dr_A}{d\mu_\mathrm{tot,A}}\frac{d\mu_\mathrm{tot,A}}{d\theta} \\
&= \left(\frac{dr_A}{d\mu_\mathrm{tot,A}} \right) \left(\frac{d\mu_\theta}{d\theta} + \frac{\tau_E}{p_0}\left(p_{AA}\frac{dr_A}{d\theta} + (M-1)p_{AB}  \frac{dr_B}{d\theta}\right) + \frac{\gamma \tau_I}{p_0} p_{EI}^* \frac{dr_I}{d\theta} \right) \\
&= \frac{\frac{dr_A}{d\mu_\mathrm{tot,A}}}{\left( 1 - \frac{dr_A}{d\mu_\mathrm{tot,A}} \frac{\tau_E}{p_0}p_{AA} \right)} \left(\frac{d\mu_\theta}{d\theta} + \frac{\tau_E}{p_0} (M-1)p_{AB}  \frac{dr_B}{d\theta} + \frac{\gamma \tau_I}{p_0} p_{EI}^* \frac{dr_I}{d\theta} \right)
\end{aligned} \end{equation}

\begin{equation} \begin{aligned}
\frac{dr_B}{d\theta} &= \frac{dr_B}{d\mu_\mathrm{tot,B}}\frac{d\mu_\mathrm{tot,B}}{d\theta} \\
&= \frac{dr_B}{d\mu_\mathrm{tot,B}} \left( \frac{\tau_E}{p_0}\left(p_{AA}\frac{dr_B}{d\theta} +  p_{AB}\frac{dr_A}{d\theta} + (M-2)p_{AB}  \frac{dr_B}{d\theta}\right) + \frac{\gamma \tau_I}{p_0} p_{EI}^* \frac{dr_I}{d\theta} \right) \\
&= \frac{\frac{dr_B}{d\mu_\mathrm{tot,B}}}{\left(1 -  \frac{dr_B}{d\mu_\mathrm{tot,B}}\frac{\tau_E}{p_0}\left(p_{AA} + (M-2)p_{AB} \right) \right)} \left( \frac{\tau_E}{p_0}  p_{AB}\frac{dr_A}{d\theta} + \frac{\gamma \tau_I}{p_0} p_{EI}^* \frac{dr_I}{d\theta} \right)
\end{aligned} \end{equation}

\begin{equation} \begin{aligned}
\frac{dr_I}{d\theta} &= \frac{dr_I}{d\mu_\mathrm{tot,I}}\frac{d\mu_\mathrm{tot,I}}{d\theta} \\
&= \frac{dr_I}{d\mu_\mathrm{tot,I}} \left( \frac{\tau_E}{p_0}p_{EI}^*\left(\frac{dr_A}{d\theta}+(M-1)\frac{dr_B}{d\theta} \right) + \frac{\gamma \tau_I}{p_0} p_{II} \frac{dr_I}{d\theta} \right) \\
&= \frac{\frac{dr_I}{d\mu_\mathrm{tot,I}}}{1 -  \frac{dr_I}{d\mu_\mathrm{tot,I}}\frac{\gamma \tau_I}{p_0} p_{II}} \left( \frac{\tau_E}{p_0}p_{EI}^*\left(\frac{dr_A}{d\theta}+(M-1)\frac{dr_B}{d\theta} \right) \right).
\end{aligned} \end{equation}

\paragraph{}
Inserting $dr_I/d\theta$ into $dr_Bd\theta$:
\begin{equation} \begin{aligned} 
\frac{dr_B}{d\theta} &= \frac{\frac{dr_B}{d\mu_\mathrm{tot,B}}}{\left(1 -  \frac{dr_B}{d\mu_\mathrm{tot,B}}\frac{\tau_E}{p_0}\left(p_{AA} + (M-2)p_{AB} \right) \right)} \left( \frac{\tau_E}{p_0}  p_{AB}\frac{dr_A}{d\theta} + \frac{\gamma \tau_I}{p_0} p_{EI}^* \frac{\frac{dr_I}{d\mu_\mathrm{tot,I}}}{1 - \frac{dr_I}{d\mu_\mathrm{tot,I}}\frac{\gamma \tau_I}{p_0} p_{II}} \left( \frac{\tau_E}{p_0}p_{EI}^*\left(\frac{dr_A}{d\theta}+(M-1)\frac{dr_B}{d\theta} \right) \right) \right) \\
 \frac{dr_B}{d\theta} &= \frac{dr_A}{d\theta} \underbrace{\frac{\frac{dr_B}{d\mu_\mathrm{tot,B}} \left( \frac{\tau_E}{p_0}  p_{AB} + \frac{\gamma \tau_I}{p_0} p_{EI}^* \left(\frac{\frac{dr_I}{d\mu_\mathrm{tot,I}}}{1 - \frac{dr_I}{d\mu_\mathrm{tot,I}} \frac{\gamma \tau_I}{p_0} p_{II}}\right)  \frac{\tau_E}{p_0}p_{EI}^* \right) }{\left(1 -  \frac{dr_B}{d\mu_\mathrm{tot,B}}\frac{\tau_E}{p_0}\left(p_{AA} + (M-2)p_{AB} \right) - \frac{\gamma \tau_I}{p_0}p_{EI}^*\frac{\frac{dr_I}{d\mu_\mathrm{tot,I}}}{1 -  \frac{dr_I}{d\mu_\mathrm{tot,I}}\frac{\gamma \tau_I}{p_0} p_{II}} \frac{\tau_E}{p_0}p_{EI}^*(M-1) \right)}}_{\equiv X} 
\end{aligned} \end{equation}
and inserting $dr_B/d\theta$ into $dr_I/d\theta$:
\begin{equation} \begin{aligned}
\frac{dr_I}{d\theta} &= \frac{dr_A}{d\theta} \underbrace{\frac{\frac{dr_I}{d\mu_\mathrm{tot,I}}}{1 -  \frac{dr_I}{d\mu_\mathrm{tot,I}}\frac{\gamma \tau_I}{p_0} p_{II}} \left( \frac{\tau_E}{p_0}p_{EI}^*\left(1+(M-1)X \right) \right)}_{\equiv Y} 
\end{aligned} \end{equation}
which yields:
\begin{equation} \begin{aligned}
\frac{dr_A}{d\theta} &= \frac{\frac{dr_A}{d\mu_\mathrm{tot,A}}}{\left( 1 - \frac{dr_A}{d\mu_\mathrm{tot,A}} \frac{\tau_E}{p_0}p_{AA} \right)} \left(\frac{d\mu_\theta}{d\theta} + \frac{\tau_E}{p_0} (M-1)p_{AB} X \frac{dr_A}{d\theta} + \frac{\gamma \tau_I}{p_0} p_{EI}^* Y\frac{dr_A}{d\theta} \right) \\
\frac{dr_A}{d\theta} &= \frac{\frac{dr_A}{d\mu_\mathrm{tot,A}}\frac{d\mu_\theta}{d\theta}}{\left( 1 - \frac{dr_A}{d\mu_\mathrm{tot,A}} \frac{\tau_E}{p_0}p_{AA}  - \frac{\tau_E}{p_0} (M-1)p_{AB} X - \frac{\gamma \tau_I}{p_0} p_{EI}^* Y\right)}.
\end{aligned} \end{equation}

% Do NOT remove this, even if you are not including acknowledgments.
\section{Acknowledgments}
We thank Ken Miller, Bard Ermentrout, Jon Rubin, Anne-Marie Oswald and Thanos Tzounopoulos for their valuable comments.

%\section{References}
\bibliography{diss_bib.bib}

%%%%%%%%%%%%%%%%%%%%%%%%%%%%%%%%%%%%%%%%%%%%%%%%%
\clearpage
\begin{figure}[ht!]
\centering \includegraphics{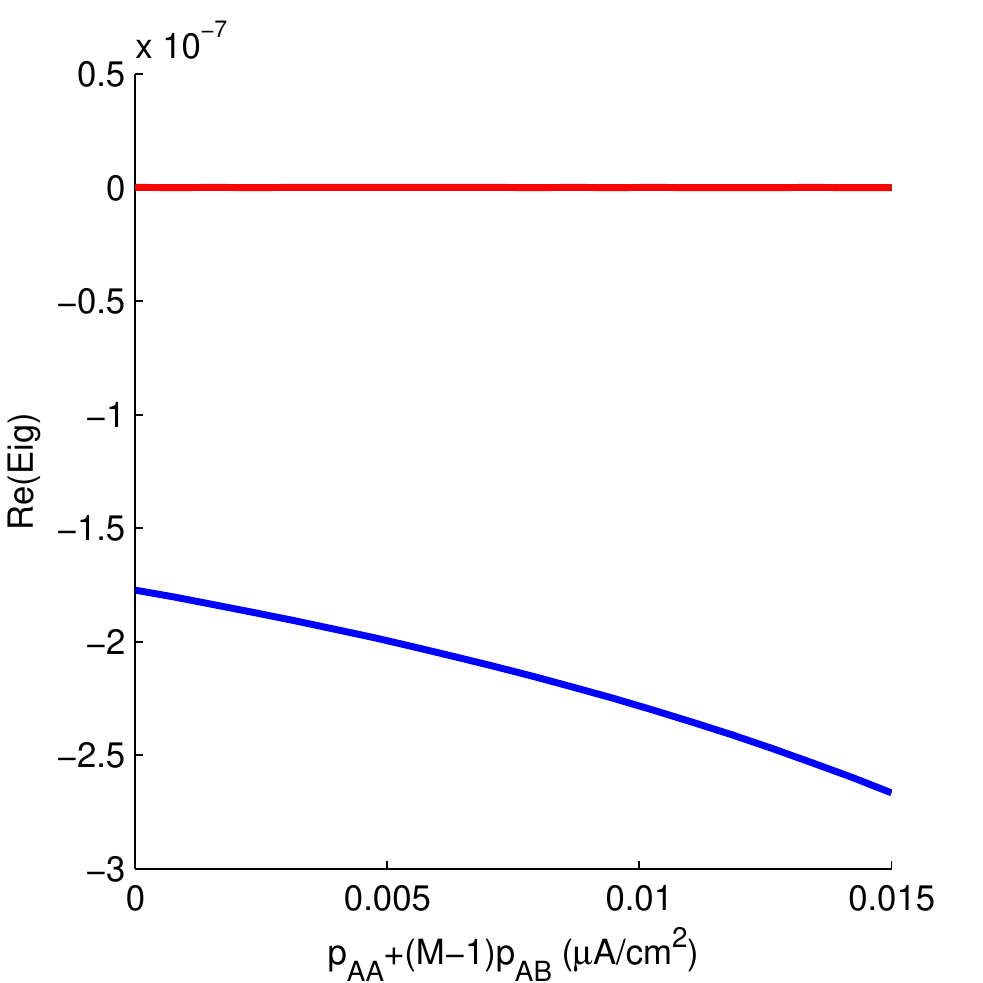}
\caption{ \bf \small
Supplemental Figure: Eigenvalues of the system Eqs. \eqref{dpEIdt_unbal}-\eqref{drIdt_unbal} with unbalanced iSTDP.
}
\end{figure}

\end{document}